\documentclass[CCIRM,XML,chapters,Unicode,screen]{cedram}
\usepackage{hyperref}
\usepackage[nohints]{minitoc}
\usepackage{graphicx}
\usepackage{color}
\usepackage{bm} 
\usepackage{amsmath}
\usepackage{amssymb} 
\usepackage{amsthm}  
\usepackage{mathrsfs} 
\usepackage{ifthen}
\usepackage{listings}
\usepackage{tikz}
\usepackage[framemethod=TikZ]{mdframed} 
\usepackage{fancyvrb} 

\usetikzlibrary{calc,positioning,shapes,arrows,chains}

\hypersetup{pdftitle={Symbolic tensor calculus on manifolds},
  pdfsubject={tensor calculus},
  pdfauthor={Eric Gourgoulhon <eric.gourgoulhon@obspm.fr>},
  pdfkeywords={LaTeX},
  colorlinks=true
}
\definecolor{dkgreen}{rgb}{0,0.6,0}
\definecolor{gray}{rgb}{0.5,0.5,0.5}
\definecolor{mauve}{rgb}{0.58,0,0.82}

\definecolor{inblue}{rgb}{0,0,0.7}
\definecolor{outred}{rgb}{0.8,0,0.0}

\lstdefinestyle{NBinput}{language=Python,
        keywordstyle=\bf\color{dkgreen},
        commentstyle=\color{blue},
        stringstyle=\color{mauve},
        morekeywords={False,True},
        showspaces=false,
        showstringspaces=false,
        fancyvrb=true,
        basewidth=0.5em,
        basicstyle=\ttfamily,
        }

\newsavebox{\FVerbBox}
\newcounter{NBin}
\newenvironment{NBin}{\refstepcounter{NBin}
  \VerbatimEnvironment
  \lstset{style=NBinput}
  \begin{lrbox}{\FVerbBox}
  \begin{minipage}[t]{5.45in}
  \begin{Verbatim}
}{
  \end{Verbatim}
  \end{minipage}
  \end{lrbox}
  \medskip
  \par\noindent {\color{inblue} \texttt{In [\theNBin]\!:}}\;\fcolorbox{gray!40}{gray!7}{\usebox{\FVerbBox}}
}

\newenvironment{NBout}{
{
  \par\noindent \color{outred} \texttt{Out[\theNBin]\!:}}
  \begin{minipage}[t]{5.5in}
} {
  \end{minipage}
  \ \\
}

\newenvironment{NBoutM}{
{
  \par\noindent \color{outred} \texttt{Out[\theNBin]\!:}}
  \begin{minipage}[t]{5.5in}
  $\displaystyle
} {
  $
  \end{minipage}
  \ \\
}

\newenvironment{NBprint}{
\par\hspace{1cm}
\BVerbatim
} {
\endBVerbatim
\ \\[1ex]
}

\newcommand{\soft}[1]{\textsf{#1}}
\newcommand{\code}[1]{\texttt{#1}}
\newcommand{\Sage}{\soft{SageMath}}
\newcommand{\SM}{\soft{SageManifolds}}
\newcommand{\footurl}[1]{\footnote{\scriptsize\url{#1}}}

\newcommand{\M}{M}
\newcommand{\R}{\mathbb{R}}
\newcommand{\K}{\mathbb{K}}
\newcommand{\Sp}{\mathbb{S}^2}
\newcommand{\X}{\mathfrak{X}}

\newcommand{\D}{\mathrm{d}}

\newcommand{\w}[1]{\bm{#1}}

\newcommand{\be}{\begin{equation}}
\newcommand{\ee}{\end{equation}}
\newcommand{\bea}{\begin{eqnarray}}
\newcommand{\eea}{\end{eqnarray}}

\newcommand{\dert}[2]{{\partial #1}/{\partial #2}}

\newcommand{\defin}[1]{\textbf{\itshape #1}}

\newcounter{remarkCounter}[section]
\newenvironment{remark}%
{\refstepcounter{remarkCounter}
\par\medskip\noindent\small\textbf{Remark \theremarkCounter:}}%
{\par\medskip}

\begin{document}

\tikzset{
base/.style={draw, thick, align=center},
native/.style={base, fill=cyan!30},
alg/.style={base, fill=red!40, rounded corners},
diff/.style = {base, fill=yellow!50, rounded corners},
dict/.style={base, fill=pink!40, draw=red},
tens/.style = {base, fill=yellow!25, align=left},
empty/.style={align=left},
legend/.style = {minimum width=2em, minimum height=1em},
native_legend/.style = {native, legend},
alg_legend/.style = {alg, legend},
diff_legend/.style = {diff, legend},
line/.style = {->, draw, thick, >=triangle 45}
}

\title{Symbolic tensor calculus on manifolds: a SageMath implementation}
\author{\firstname{Éric} \lastname{Gourgoulhon}}
\address{Laboratoire Univers et Théories \\
CNRS, Observatoire de Paris, Université Paris Diderot,
Université Paris Sciences et Lettres \\
92190 Meudon, France}
\email{eric.gourgoulhon@obspm.fr}
\email{marco.mancini@obspm.fr}
\author{\firstname{Marco} \lastname{Mancini}}

\maketitle

\setcounter{minitocdepth}{1}
\dominitoc

\section*{Preface}
These notes correspond to two lectures given by one of us (EG)
at \emph{Journées Nationales de Calcul Formel 2018}  (French Computer Algebra Days),
which took place at Centre International de Rencontres Mathématiques (CIRM),
in Marseille, France, on 22-26 January 2018.
The slides, demo
notebooks and videos of these lectures are available at
\begin{center}
\url{https://sagemanifolds.obspm.fr/jncf2018/}
\end{center}

EG warmly thanks Marc Mezzarobba and the organizers of JNCF 2018 for their
invitation and the perfect organization of the lectures. He also acknowledges
the great hospitality of CIRM and many fruitful exchanges with the conference
participants. We are very grateful to Travis Scrimshaw for his help in the
writing of these notes, especially for providing a customized \LaTeX{} environment
to display Jupyter notebook cells.

\newpage
\tableofcontents

\chapter{Introduction} \label{s:int}

\minitoc

\section{What is tensor calculus on manifolds?}

We shall provide precise definitions in
Chaps.~\ref{s:man} and~\ref{s:vec}. Here, let us state briefly that
\emph{tensor calculus on manifolds} stands for calculus on vector fields,
and more generally tensor fields, on differentiable manifolds, involving the following operations~\cite{Lee13}:
\begin{itemize}
\item arithmetics of tensor fields;
\item tensor product, contraction;
\item (anti)symmetrization;
\item Lie derivation along vector fields;
\item pullback and pushforward associated with smooth manifold maps;
\item exterior (Cartan) calculus on differential forms;
\item covariant derivation with respect to a given affine connection;
\item evaluating the torsion and the curvature of an affine connection.
\end{itemize}
Moreover, on pseudo-Riemannian manifolds, i.e. differentiable manifolds endowed
with a metric tensor, we may add the following
operations~\cite{Lee97,ONeil83}:
\begin{itemize}
\item musical isomorphisms (i.e. raising and lowering indices with the metric tensor);
\item determining the Levi-Civita connection;
\item evaluating the curvature tensor of the Levi-Civita connection (Riemann tensor);
\item Hodge duality;
\item computing geodesics.
\end{itemize}

\section{A few words of history}

Symbolic tensor calculus has a long history, which started
almost as soon as computer algebra itself in the 1960s.
Probably, the first tensor calculus program was \soft{GEOM}, written by J.G.~Fletcher
in 1965~\cite{Fletc67}. Its main capability was to compute the Riemann tensor
of a given metric. In 1969, R.A.~d'Inverno developed \soft{ALAM}
(for \emph{Atlas Lisp Algebraic Manipulator}) and used it to compute
the Riemann and Ricci tensors of the Bondi metric.
According to~\cite{Skea94},
the original calculations took Bondi and collaborators 6 months to finish,
while the computation with \soft{ALAM} took 4 minutes and yielded the
discovery of 6 errors in the original paper by Bondi et al.
Since then, numerous packages have been developed; the reader is referred to~\cite{MacCa18}
for a recent review of computer algebra
systems for general relativity (see also \cite{MacCa02} for a review up to 2002), and to~\cite{KorolKS13,BirkaGSC17} for more recent reviews
focused on tensor calculus.
It is also worth to point out the extensive list of
tensor calculus packages maintained by J. M. Martin-Garcia at
\url{http://www.xact.es/links.html}.


\section{Software for differential geometry}

Software packages for differential geometry and tensor calculus can be
classified in two categories:
\begin{enumerate}
\item Applications atop some general purpose computer algebra system.
Notable examples\footnote{See \url{https://en.wikipedia.org/wiki/Tensor_software}
for more examples.} are
the \soft{xAct} suite~\cite{Marti08} and \soft{Ricci}~\cite{ricci}, both
running atop \soft{Mathematica},
\soft{DifferentialGeometry}~\cite{AnderT12} integrated into \soft{Maple},
\soft{GRTensorIII}~\cite{grtensorIII} atop \soft{Maple}, \soft{Atlas 2}~\cite{atlas2}
for \soft{Mathematica} and \soft{Maple}, \soft{ctensor} and \soft{itensor} for \soft{Maxima}~\cite{Toth05}
and
\soft{SageManifolds}~\cite{sagemanifolds} integrated in \soft{SageMath}.
\item Standalone applications. Recent examples are \soft{Cadabra}~\cite{Peete07} (field theory),
\soft{SnapPy}~\cite{snappy} (topology and geometry of 3-manifolds)  and
\soft{Redberry}~\cite{BolotP13} (tensors); older examples can be found in
Refs.~\cite{MacCa02,MacCa18}.
\end{enumerate}
All applications listed in the second category are free software. In
the first category, \soft{xAct} and \soft{Ricci} are also free software, but
they require a proprietary product, the source code of which is closed (\soft{Mathematica}).

As far as tensor calculus is concerned, the above packages can be distinguished by
the type of computation that they perform:
abstract calculus (\soft{xAct/xTensor}, \soft{Ricci}, \soft{itensor},
\soft{Cadabra}, \soft{Redberry}),
or component calculus (\soft{xAct/xCoba}, \soft{DifferentialGeometry}, \soft{GRTensorIII},
\soft{Atlas 2}, \soft{ctensor}, \soft{SageManifolds}).
In the first category, tensor operations such as contraction or covariant differentiation
are performed by manipulating the indices themselves rather than the components
to which they correspond. In the second category, vector frames are explicitly
introduced on the manifold and tensor operations are carried out on the components
in a given frame.


\section{A brief overview of SageMath} \label{s:int:overview_Sage}

Since the tensor calculus method presented here is implemented in \Sage{}, we
give first a brief overview of the latter.

\Sage{}\footnote{\url{http://www.sagemath.org}} is a free, open-source mathematics software system, which is
based on the Python programming language. It makes use of over 90 open-source packages,
among which are \soft{Maxima}, \soft{Pynac} and \soft{SymPy} (symbolic calculations),
\soft{GAP} (group theory),
\soft{PARI/GP} (number theory), \soft{Singular} (polynomial computations),
\soft{matplotlib} (high quality 2D figures), and \soft{Jupyter} (graphical interface).
\Sage{} provides a uniform Python interface to all these packages; however,
\Sage{} is much more than a mere interface: it contains a large and increasing part of
original code (more than 750,000 lines of Python and Cython, involving 5344 classes).
\Sage{} was created in 2005 by William Stein~\cite{SteinJ05} and since
then its development has been sustained by more than a hundred researchers
(mostly mathematicians). In particular, a strong impulse is currently being
provided by the European Horizon 2020 project OpenDreamKit~\cite{OpenDreamKit}.
Very good introductory textbooks about \Sage{} are
\cite{JoyneS14,Zimme13,Zimme18,Bard15}.

Apart from the syntax, which is based on a popular programming language
(Python) and not on some custom script
language, a difference between \Sage{} and, e.g., \soft{Maple} or \soft{Mathematica}
is the use of the \emph{parent/element pattern}\index{parent}\index{element}. This pattern closely
reflects actual mathematics.
For instance, in \soft{Mathematica}, all objects
are trees of symbols and the program is essentially a set of
sophisticated rules to manipulate symbols. On the contrary, in \Sage{}
each object has a given type (i.e. is an instance of a given
Python class\footnote{Let us
recall that within an object-oriented programming language (as Python),
a \defin{class} is a structure to declare and store the
properties common to a set of objects. These properties
are data (called
\defin{attributes} or \defin{state variables}) and functions acting
on the data (called \defin{methods}). A specific realization of an object
within a given class is called an \defin{instance} of that class.}),
and one distinguishes \defin{parent} types, which model mathematical
sets with some structure (e.g. algebraic structure), from \defin{element} types,
which model set elements. Moreover, each parent belongs to some
dynamically generated class that encodes information
about its \emph{category}, in the mathematical sense of the word.\footnote{See
\url{http://doc.sagemath.org/html/en/reference/categories/sage/categories/primer.html}
for a discussion of \Sage{}'s category framework}
Automatic conversion rules, called \defin{coercions}\index{coercion},
prior to a binary operation, e.g. $x+y$ with $x$ and $y$ having different
parents, are implemented.

\section{The purpose of this lecture}

This lecture aims at presenting
a \emph{symbolic tensor calculus method} that
\begin{itemize}
\item runs on fully specified smooth manifolds (described by an atlas);
\item is not limited to a single coordinate chart or vector frame;
\item runs even on non-parallelizable manifolds (i.e.\ manifolds that cannot
be covered by a single vector frame);
\item is independent of the symbolic backend (e.g., \soft{Pynac/Maxima},
\soft{SymPy}, ...) used to perform calculus at the level of coordinate expressions.
\end{itemize}
The aim is to present not only the main ideas of the method, but also some details of its implementation in \Sage{}. This implementation has been
performed via the \soft{SageManifolds} project:
\begin{center}
\url{https://sagemanifolds.obspm.fr},
\end{center}
the contributors to which are listed at
\begin{center}
\url{https://sagemanifolds.obspm.fr/authors.html}.
\end{center}

\chapter{Differentiable manifolds} \label{s:man}

\minitoc

\section{Introduction}

Starting from basic mathematical definitions, we present
the implementation of manifolds and coordinate charts in \Sage{}
(Sec.~\ref{s:bas:manif}).
We then focus on the algebra of scalar fields on a manifold~\eqref{s:man:scalar_field_algebra}.
As we shall see in Chap.~\ref{s:vec}, this algebra plays a central role in the implementation of vector fields, the latter being considered as forming a module
over it.

\section{Differentiable manifolds} \label{s:bas:manif}

\subsection{Topological manifolds} \label{s:bas:def_manif}

Let $\K$ be a topological field. In most applications $\K=\R$ or $\K=\mathbb{C}$.
Given an integer $n\geq 1$, a \defin{topological manifold of dimension $n$ over $\K$}\index{manifold}\index{dimension of a manifold} is a topological space $\M$ obeying the following properties:
\begin{enumerate}
\item $\M$ is a \defin{separated space}\index{separated space} (also called \defin{Hausdorff space}\index{Hausdorff space}): any two distinct points of $\M$
admit disjoint open neighbourhoods.
\item $\M$ has a \defin{countable base}\index{countable base}:\footnote{In the language of topology, one says that $\M$ is a \emph{second-countable space}.}
there exists a countable family
$(U_k)_{k\in\mathbb{N}}$ of open sets of $\M$ such that any open set of $\M$ can be written as the union (possibly infinite) of some members of this family.
\item Around each point of $\M$, there exists a neighbourhood which is
homeomorphic to an open subset of $\K^n$.
\end{enumerate}
Property 1 excludes manifolds with ``forks''.
Property~2 excludes ``too large'' manifolds; in particular it permits setting
up the theory of integration on manifolds. In the case $\K=\R$, it also
allows for a smooth manifold of dimension $n$ to be embedded smoothly into the Euclidean space $\R^{2n}$
(Whitney theorem\index{Whitney theorem}).
Property~3 expresses the essence of a manifold: it means that, locally,
$\M$ ``resembles'' $\K^n$.

Let us start to discuss the implementation of manifolds in \Sage{}. We shall
do it on a concrete example, exposed in a Jupyter notebook which can be downloaded
from the page devoted to these lectures:
\begin{center}
\url{https://sagemanifolds.obspm.fr/jncf2018/}
\end{center}
As for all \Sage{}, the syntax used in this notebook is \soft{Python} one. However, no
a priori knowledge of \soft{Python} is required, since we shall explain the
main notations as they appear. The first cell of the Jupyter notebook is to have
all outputs rendered with \LaTeX:
\begin{NBin}
%display latex
\end{NBin}
\ \\
In \Sage{}, manifolds are constructed by means of the global function \code{Manifold}:
\begin{NBin}
M = Manifold(2, 'M')
print(M)
\end{NBin}
\begin{NBprint}
2-dimensional differentiable manifold M
\end{NBprint}
By default, the function \code{Manifold} returns a manifold over $\K=\R$:
\begin{NBin}
M.base_field()
\end{NBin}
\begin{NBoutM}
\mathbf{R}
\end{NBoutM}
Note the use of the standard object-oriented notation (ubiquitous in
\soft{Python}): the method \code{base\_field()} is called on the object \code{M};
since this method does not require any extra argument (all the information lies
in \code{M}), its argument list is empty, hence the final \code{()}.
Base fields different from $\mathbb{R}$
must be specified with the optional keyword \code{field}, for instance
\begin{flushleft}
\code{M = Manifold(2, 'M', field='complex')}
\end{flushleft}
to construct a complex manifold\footnote{Note however that the functionalities
regarding complex manifolds are pretty limited at the moment. Volunteers are
welcome to implement them! See \url{https://sagemanifolds.obspm.fr/contrib.html}.}.
We may check that $M$ is a topological space:
\begin{NBin}
M in Sets().Topological()
\end{NBin}
\begin{NBout}
\texttt{True}
\end{NBout}
Actually, $M$ belongs to the following categories:
\begin{NBin}
M.categories()
\end{NBin}
\begin{NBout}
\vspace{-25pt}  
\begin{gather*}
\left[\mathbf{Smooth}_{\mathbf{R}}, \mathbf{Differentiable}_{\mathbf{R}}, \mathbf{Manifolds}_{\mathbf{R}}, \mathbf{TopologicalSpaces}(\mathbf{Sets}), \right.
\\ \left. \mathbf{Sets},  \mathbf{SetsWithPartialMaps}, \mathbf{Objects}\right]
\end{gather*}
\end{NBout}
As we can see from the first category in the above list, \code{Manifold}
constructs a smooth manifold by default.
If one would like to stick to the topological level, one should add
the keyword argument \code{structure='topological'} to \code{Manifold},
i.e.
\begin{flushleft}
\code{M = Manifold(2, 'M', structure='topological')}
\end{flushleft}
Then $M$ would have been a topological manifold without any further structure.

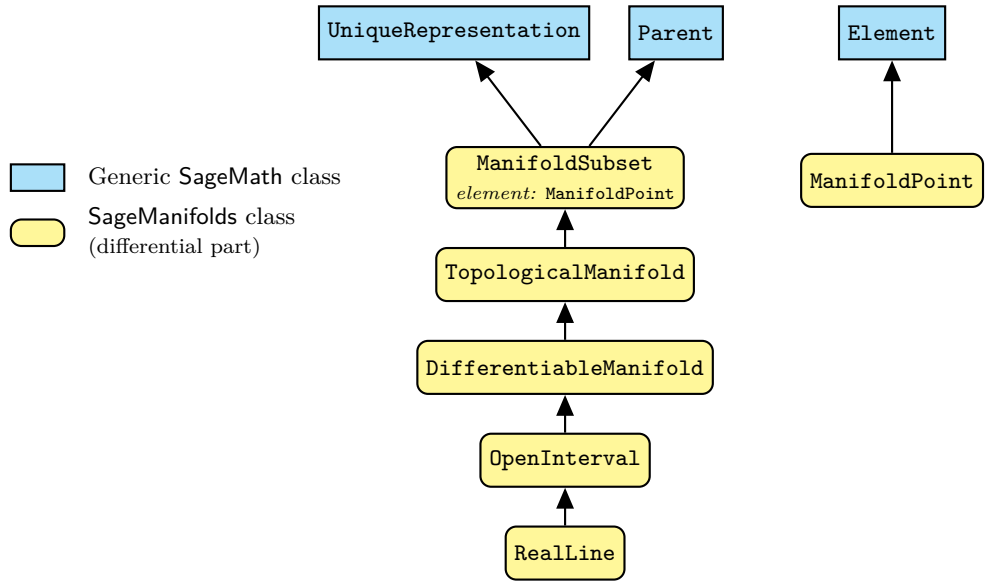
\begin{figure}
\begin{center}
\begin{tikzpicture}[font=\small, node distance=0.5cm, minimum
height=2em, auto]

\node[native](unique_representation)
{\code{UniqueRepresentation}};

\node[native, right=of unique_representation](parent)
{\code{Parent}};

\coordinate (Middle) at ($(unique_representation)!0.5!(parent)$);

\node[diff, below=1.5cm of Middle](subset)
{\code{ManifoldSubset}\\ {\scriptsize {\it element:} \code{ManifoldPoint}}};
\path[line] (subset) -- (unique_representation);
\path[line] (subset) -- (parent);

\node[diff, below=of subset](manifold)
{\code{TopologicalManifold}};
\path[line] (manifold) -- (subset);

\node[diff, below=of manifold](diffmanifold)
{\code{DifferentiableManifold}};
\path[line] (diffmanifold) -- (manifold);

\node[diff, below=of diffmanifold](interval)
{\code{OpenInterval}};
\path[line] (interval) -- (diffmanifold);

\node[diff, below=of interval](realline)
{\code{RealLine}};
\path[line] (realline) -- (interval);

\node[native, right=1.5cm of parent](element)
{\code{Element}};

\node[diff, below=1.225cm of element](point)
{\code{ManifoldPoint}};

\path[line] (point) -- (element);


\node[native_legend, left=5cm of subset]
(native_legend){};
\node[empty, right=0.5em of native_legend]
{Generic \Sage{} class};

\node[diff_legend, below=1.em of native_legend]
(diff_legend){};
\node[empty, right=0.5em of diff_legend]
{\soft{SageManifolds} class\\ \footnotesize (differential part)};

\end{tikzpicture}
\end{center}
\caption{\label{f:man:domain_classes}\footnotesize
Python classes for topological
manifolds, differentiable manifolds, subsets of them
and points on them (\code{ManifoldPoint}).}
\end{figure}

Manifolds are implemented by the Python classes \code{TopologicalManifold}
and \\ \code{DifferentiableManifold} (see Fig.~\ref{f:man:domain_classes}),
actually by dynamically generated subclasses of those, via \Sage{} category
framework:\footnote{See \url{http://doc.sagemath.org/html/en/reference/categories/sage/categories/primer.html}\index{category} for details.}
\begin{NBin}
type(M)
\end{NBin}
\begin{NBout}
\begin{verbatim}
<class 'sage.manifolds.differentiable.manifold.
        DifferentiableManifold_with_category'>
\end{verbatim}
\end{NBout}
Let us check that the actual class of \code{M}, i.e.\ \code{DifferentiableManifold-with-category},
is a subclass of \code{DifferentiableManifold}:
\begin{NBin}
isinstance(M,
           sage.manifolds.differentiable.manifold.DifferentiableManifold)
\end{NBin}
\begin{NBout}
\texttt{True}
\end{NBout}
and hence of
\code{TopologicalManifold} according to the inheritance diagram of Fig.~\ref{f:man:domain_classes}:
\begin{NBin}
isinstance(M, sage.manifolds.manifold.TopologicalManifold)
\end{NBin}
\begin{NBout}
\texttt{True}
\end{NBout}
Notice from Fig.~\ref{f:man:domain_classes} that
\code{TopologicalManifold} itself is a subclass of \code{ManifoldSubset} (the class
for generic subsets of a manifold), which reflects the fact that $M\subset M$.

\subsection{Coordinate charts} \label{s:man:coord_chart}

Property~3 in the definition of a topological manifold (Sec.~\ref{s:bas:def_manif})
means that one can label the points of $\M$ in a
continuous way by $n$ numbers $(x^\alpha)_{\alpha\in\{0,\ldots,n-1\}}\in \K^n$,
which are called \defin{coordinates}\index{coordinate}.
More precisely, given an open subset $U\subset\M$, a \defin{coordinate chart}\index{coordinate!chart}
(or simply a \defin{chart}\index{chart})
on $U$ is a homeomorphism\footnote{Let us recall that a  \defin{homeomorphism}\index{homeomorphism} between two topological spaces
(here $U$ and $X(U)$) is a bijective map $X$ such
that both $X$ and $X^{-1}$ are continuous.}
\be \label{e:man:def_chart}
    \begin{array}{rccl}
    X: & U\subset \M & \longrightarrow & X(U)\subset \K^n \\
        & p & \longmapsto & (x^0, \ldots, x^{n-1}) .
    \end{array}
\ee
We declare a chart, along with the symbols used to denote the coordinates
(here $x=x^0$ and $y=x^1$) by
\begin{NBin}
U = M.open_subset('U')
XU.<x,y> = U.chart()
XU
\end{NBin}
\begin{NBoutM}
(U, (x,y))
\end{NBoutM}
Open subsets of a differentiable manifold are implemented by a (dynamically generated) subclass of
\code{DifferentiableManifold}, since they are differentiable manifolds in their own:
\begin{NBin}
isinstance(U,
           sage.manifolds.differentiable.manifold.DifferentiableManifold)
\end{NBin}
\begin{NBout}
\texttt{True}
\end{NBout}

Points on $M$ are created from their coordinates in a given chart:
\begin{NBin}
p = U((1,2), chart=XU, name='p')
print(p)
\end{NBin}
\begin{NBprint}
Point p on the 2-dimensional differentiable manifold M
\end{NBprint}
The syntax \code{U(...)} used to create $p$ as an element of $U$
reflects the parent/element\index{parent}\index{element} pattern employed in \Sage{}; indeed $U$
is the parent of $p$:
\begin{NBin}
p.parent()
\end{NBin}
\begin{NBoutM}
U
\end{NBoutM}
Points are implemented by a dynamically generated subclass of
\code{ManifoldPoint} (cf.\ Fig.~\ref{f:man:domain_classes}).
The principal attribute of this class is the one storing the point's coordinates
in various charts; it is implemented as
a Python dictionary,\footnote{A \defin{dictionary}\index{dictionary},
also known as \defin{associative array}, is a
data structure that generalizes the concept of array in the sense that the
key to access to an element is not restricted to an integer or a tuple of integers.} whose keys are the charts:
\begin{NBin}
p._coordinates
\end{NBin}
\begin{NBoutM}
\{(U, (x,y)): (1,2)\}
\end{NBoutM}
The leading underscore in the name \code{\_coordinates} is a notation
convention to specify that this attribute is a \defin{private} one:
the dictionary \code{\_coordinates} should
not be manipulated by the end user or involved in some code
outside of the class \code{ManifoldPoint}.
It belongs to the internal implementation, which may be changed while
the user interface of the class \code{ManifoldPoint} is kept fixed. We show
this private attribute here
because we are precisely interested in implementation features.
The public way to recover the point's coordinates is to let the chart act on
the point (reflecting thereby the definition~\eqref{e:man:def_chart} of a chart):
\begin{NBin}
XU(p)
\end{NBin}
\begin{NBoutM}
(1,2)
\end{NBoutM}

Usually, one needs more than a single coordinate system to cover $\M$.
An \defin{atlas}\index{atlas} on $\M$ is a set of pairs
$(U_i,X_i)_{i\in I}$, where $I$ is a set, $U_i$ an open set of $\M$ and $X_i$ a chart on $U_i$,
such that the union of all $U_i$'s covers $\M$:
\be
    \bigcup_{i\in I} U_i = \M.
\ee
Here we introduce a second chart on $\M$:
\begin{NBin}
V = M.open_subset('V')
XV.<xp,yp> = V.chart("xp:x' yp:y'")
XV
\end{NBin}
\begin{NBoutM}
\left( V, (x', y') \right)
\end{NBoutM}
and declare that $\M$ is covered by only two charts, i.e.\ that $M=U\cup V$:
\begin{NBin}
M.declare_union(U, V)
\end{NBin}
\begin{NBin}
M.atlas()
\end{NBin}
\begin{NBoutM}
\left[ \left(U, (x,y) \right), \left( V, (x', y') \right) \right]
\end{NBoutM}

\subsection{Smooth manifolds}

For manifolds, the concept of differentiability is
defined from the smooth structure of $\K^n$, via an atlas:
a \defin{smooth manifold}\index{smooth!manifold}\index{manifold!smooth --},
is a topological manifold $\M$ equipped with an atlas
$(U_i,X_i)_{i\in I}$ such that for any non-empty intersection
$U_i \cap U_j$, the map
\be \label{e:bas:transition_map}
    X_i \circ X_j^{-1} : X_j(U_i \cap U_j)
    \subset \K^n \longrightarrow X_i(U_i \cap U_j)
    \subset \K^n
\ee
is smooth (i.e.~$C^\infty$).
Note that the above map is from an open set of $\K^n$ to an open set of $\K^n$, so that the invoked differentiability is nothing but that of $\K^n$.
Such a map is called a \defin{change of coordinates}\index{change!of coordinates}\index{coordinate!change} or, in the mathematical literature, a
\defin{transition map}\index{transition map}.
The atlas $(U_i,X_i)_{i\in I}$ is called a
\defin{smooth atlas}\index{smooth!atlas}\index{atlas!smooth --}.

\begin{remark}
Strictly speaking a smooth manifold is a pair $(\M,\mathcal{A})$  where
$\mathcal{A}$ is a (maximal) smooth atlas on $\M$.
Indeed a given topological manifold $\M$
can have non-equivalent differentiable structures, as shown by Milnor (1956)~\cite{Milno56}
in the specific case of the unit sphere of dimension~7, $\mathbb{S}^7$: there exist smooth manifolds, the so-called \emph{exotic spheres}\index{exotic!sphere},
that are homeomorphic to $\mathbb{S}^7$ but not diffeomorphic
to $\mathbb{S}^7$.  On the other side, for $n\leq 6$, there is a unique smooth
structure for the sphere $\mathbb{S}^n$.
Moreover, any manifold of dimension $n\leq 3$ admits a unique smooth structure.
Amazingly, in the case of $\R^n$, there exists a unique smooth structure (the standard one) for any $n\not=4$, but for $n=4$ (the spacetime case!) there exist uncountably many non-equivalent smooth structures, the so-called
\emph{exotic $\R^4$}\index{exotic!$\R^4$}~\cite{Taube87}.
\end{remark}

For the manifold $M$ under consideration, we define the transition map \code{XU}~$\to$~\code{XV}
on $W = U\cap V$ as follows:
\begin{NBin}
XU_to_XV = XU.transition_map(XV,
                             (x/(x^2+y^2), y/(x^2+y^2)),
                             intersection_name='W',
                             restrictions1= x^2+y^2!=0,
                             restrictions2= xp^2+yp^2!=0)
XU_to_XV.display()
\end{NBin}
\begin{NBoutM}
\begin{cases}
x' = \frac{x}{x^2+y^2} \\
y' = \frac{y}{x^2+y^2}
\end{cases}
\end{NBoutM}
The argument \code{restrictions1} means that
$W = U\setminus \{S\}$, where $S$ is the point of coordinates $(x,y)=(0,0)$,
while the argument \code{restrictions2} means that
$W = V\setminus \{N\}$, where $N$ is the point of coordinates $(x',y')=(0,0)$.
Since $M=U\cup V$, we have then
\be
    U = M \setminus \{N\},\qquad
    V = M \setminus \{S\},\quad\mbox{and}\quad
    W = M \setminus \{N, S\} .
\ee
The transition map \code{XV}~$\to$~\code{XU} is obtained by computing the inverse
of the one defined above:
\begin{NBin}
XU_to_XV.inverse().display()
\end{NBin}
\begin{NBoutM}
\begin{cases}
x = \frac{x'}{x'^2+y'^2} \\
y = \frac{y'}{x'^2+y'^2}
\end{cases}
\end{NBoutM}
At this stage, the smooth manifold $M$ is fully specified, being covered by
one atlas with all transition maps specified. The reader may have recognized that
$M$ is nothing but the 2-dimensional sphere:
\be
    M = \Sp ,
\ee
with \code{XU} (resp. \code{XV}) being
the chart of \defin{stereographic coordinates}\index{stereographic!coordinates}
from the North pole $N$ (resp. the South pole $S$).

Since the transition maps have been defined,
we can ask for the coordinates $(x',y')$ of the point $p$, whose $(x,y)$
coordinates were $(1,2)$:
\begin{NBin}
XV(p)
\end{NBin}
\begin{NBoutM}
\left( \frac{1}{5}, \frac{2}{5} \right)
\end{NBoutM}
This operation has updated the internal dictionary \code{\_coordinates}
(compare with \code{Out~[13]}):
\begin{NBin}
p._coordinates
\end{NBin}
\begin{NBoutM}
\left\{ (U, (x,y)) : (1,2), \left( V, (x', y') \right) : \left( \frac{1}{5}, \frac{2}{5} \right) \right\}
\end{NBoutM}

\subsection{Smooth maps}

Given two smooth manifolds, $\M$ and $\M'$, of
respective dimensions $n$ and $n'$, we say that a map
$\Phi : \M \rightarrow \M'$ is \defin{smooth map}\index{smooth!map} if and only if in some (and hence all, thanks to the smoothness of~\eqref{e:bas:transition_map}) coordinate systems
of $\M$ and $\M'$ belonging to the smooth atlases of $\M$ and $\M'$,
the coordinates of the image $\Phi(p)$ of any point $p\in M$
are smooth functions $\K^n\rightarrow \K^{n'}$ of the coordinates of $p$.
The map $\Phi$ is said to be a \defin{diffeomorphism}\index{diffeomorphism} iff
it is bijective and both $\Phi$ and $\Phi^{-1}$ are smooth. This implies $n=n'$.

Back to our example manifold, a natural smooth map is the embedding of $\Sp$ in
$\R^3$. To define it, we start by declaring $\R^3$ as a 3-dimensional smooth
manifold, canonically endowed with a single chart, that of Cartesian coordinates
$(X,Y,Z)$:
\begin{NBin}
R3 = Manifold(3, 'R^3', r'\mathbb{R}^3')
XR3.<X,Y,Z> = R3.chart()
XR3
\end{NBin}
\begin{NBoutM}
\left( \mathbb{R}^3, (X,Y,Z) \right)
\end{NBoutM}
The embedding $\Phi: \Sp \to \R^3$ is then defined in terms of its coordinate
expression in the two charts covering $M=\Sp$:
\begin{NBin}
Phi = M.diff_map(R3, {(XU, XR3):
                      [2*x/(1+x^2+y^2), 2*y/(1+x^2+y^2),
                       (x^2+y^2-1)/(1+x^2+y^2)],
                      (XV, XR3):
                       [2*xp/(1+xp^2+yp^2), 2*yp/(1+xp^2+yp^2),
                        (1-xp^2-yp^2)/(1+xp^2+yp^2)]},
                 name='Phi', latex_name=r'\Phi')
Phi.display()
\end{NBin}
\begin{NBoutM}
\begin{array}{llcl} \Phi:& M & \longrightarrow & \mathbb{R}^3 \\ \mbox{on}\ U : & \left(x, y\right) & \longmapsto & \left(X, Y, Z\right) = \left(\frac{2 \, x}{x^{2} + y^{2} + 1}, \frac{2 \, y}{x^{2} + y^{2} + 1}, \frac{x^{2} + y^{2} - 1}{x^{2} + y^{2} + 1}\right) \\ \mbox{on}\ V : & \left({x'}, {y'}\right) & \longmapsto & \left(X, Y, Z\right) = \left(\frac{2 \, {x'}}{{x'}^{2} + {y'}^{2} + 1}, \frac{2 \, {y'}}{{x'}^{2} + {y'}^{2} + 1}, -\frac{{x'}^{2} + {y'}^{2} - 1}{{x'}^{2} + {y'}^{2} + 1}\right) \end{array}
\end{NBoutM}
We may use $\Phi$ for graphical purposes, for instance to display the grids
of the stereographic charts \code{XU} (in red) and \code{XV} (in green),
with the point $p$ atop:
\begin{NBin}
graph = XU.plot(chart=XR3, mapping=Phi, number_values=25,
                label_axes=False) + \
        XV.plot(chart=XR3, mapping=Phi, number_values=25,
                color='green', label_axes=False) + \
        p.plot(chart=XR3, mapping=Phi, label_offset=0.05)
show(graph, viewer='threejs', online=True)
\end{NBin}
\begin{center}
\includegraphics[width=0.8\textwidth]{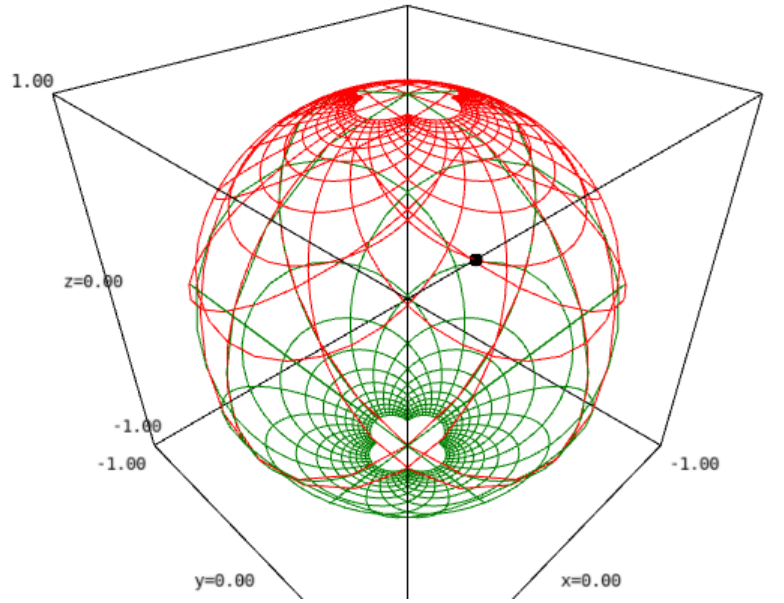}
\end{center}


\section{Scalar fields and their algebra} \label{s:man:scalar_field_algebra}

\subsection{Definition and implementation} \label{s:man:def_scalar}

Given a smooth manifold $\M$ over a topological field $\K$,
a \defin{scalar field}\index{scalar!field} (also called a
\defin{scalar-valued function}\index{scalar-valued function}) on $\M$
is a smooth map
\be
    \begin{array}{lcll}
    f: & \M &\longrightarrow & \K \\
       & p & \longmapsto  & f(p) .
    \end{array}
\ee
A scalar field has different coordinate representations $F$, $\hat F$, etc.
in different charts $X$, $\hat X$, etc. defined on $M$:
\be \label{e:man:repr_scalar_field}
    f(p) =
F(\underbrace{x^1,\ldots, x^n}_{\mbox{coord. of $p$}\atop\mbox{in chart $X$}})
= {\hat F}(\underbrace{{\hat x}^1,\ldots, {\hat x}^n}_{\mbox{coord. of $p$}\atop\mbox{in chart $\hat X$}})
= \ldots
\ee

In \Sage{}, scalar fields are implemented by the class
\code{DiffScalarField}\footurl{http://doc.sagemath.org/html/en/reference/manifolds/sage/manifolds/differentiable/scalarfield.html}
and the various representations~\eqref{e:man:repr_scalar_field} are
stored in the private attribute \code{\_express} of this class, which is a
Python dictionary
whose keys are the various charts defined on $M$:
\be \label{e:f_express}
 f.\mbox{\texttt{\_express}} = \left\{ X: F,\ \hat X: \hat F, \ldots \right\} .
\ee
Each representation $F$ is an instance of the class \code{ChartFunction},
devoted to functions of coordinates, allowing for different internal representations:
\Sage{} symbolic expression, \soft{SymPy} expression, etc.

For instance, let us define a scalar field on our example manifold $M=\Sp$:
\begin{NBin}
f = M.scalar_field({XU: 1/(1+x^2+y^2), XV: (xp^2+yp^2)/(1+xp^2+yp^2)},
                   name='f')
f.display()
\end{NBin}
\begin{NBoutM}
\begin{array}{llcl} f:& M & \longrightarrow & \mathbb{R} \\ \mbox{on}\ U : & \left(x, y\right) & \longmapsto & \frac{1}{x^{2} + y^{2} + 1} \\ \mbox{on}\ V : & \left({x'}, {y'}\right) & \longmapsto & \frac{{x'}^{2} + {y'}^{2}}{{x'}^{2} + {y'}^{2} + 1} \end{array}
\end{NBoutM}
The internal dictionary \code{\_express} is then
\begin{NBin}
f._express
\end{NBin}
\begin{NBoutM}
\left\{\left(U,(x, y)\right) : \frac{1}{x^{2} + y^{2} + 1}, \left(V,({x'}, {y'})\right) : \frac{{x'}^{2} + {y'}^{2}}{{x'}^{2} + {y'}^{2} + 1}\right\}
\end{NBoutM}
The reader may wonder about the compatibility of the two coordinate expressions
provided in the definition of $f$. Actually, to ensure the compatibility, it
is possible to declare the scalar field in a single chart, \code{XU} say,
and then to obtain its expression in chart \code{XV} by analytic continuation
from the expression in $W=U\cap V$, where both expressions are known, thanks
to the transition map \code{XV}~$\to$~\code{XU}:
\begin{NBin}
f0 = M.scalar_field({XU: 1/(1+x^2+y^2)})
f0.add_expr_by_continuation(XV, U.intersection(V))
f == f0
\end{NBin}
\begin{NBout}
\texttt{True}
\end{NBout}
The representation of the scalar field in a given chart, i.e.\ the public access
to the private directory \code{\_express}, is obtained via the method \code{coord\_function()}:
\begin{NBin}
fU = f.coord_function(XU)
fU.display()
\end{NBin}
\begin{NBoutM}
\left(x, y\right) \mapsto \frac{1}{x^{2} + y^{2} + 1}
\end{NBoutM}
\begin{NBin}
fV = f.coord_function(XV)
fV.display()
\end{NBin}
\begin{NBoutM}
\left({x'}, {y'}\right) \mapsto \frac{{x'}^{2} + {y'}^{2}}{{x'}^{2} + {y'}^{2} + 1}
\end{NBoutM}
As mentioned above, each chart representation is an instance of the
class \code{ChartFunction}:
\begin{NBin}
isinstance(fU, sage.manifolds.chart_func.ChartFunction)
\end{NBin}
\begin{NBout}
\texttt{True}
\end{NBout}
Mathematically, \defin{chart functions}\index{chart!function} are $\K$-valued functions on the codomain of
the considered chart. They map coordinates to elements of the base field $\K$:
\begin{NBin}
fU(1,2)
\end{NBin}
\begin{NBoutM}
\frac{1}{6}
\end{NBoutM}
\begin{NBin}
fU(*XU(p))
\end{NBin}
\begin{NBoutM}
\frac{1}{6}
\end{NBoutM}
Note the use of Python's star operator in \code{*XU(p)} to unpack the tuple of coordinates
returned by \code{XU(p)} (in the present case: \code{(1,2)}) to positional arguments
for the function \code{fU} (in the present case: \code{1, 2}).
On their side, scalar fields map \emph{manifold points}, not coordinates, to $\K$:
\begin{NBin}
f(p)
\end{NBin}
\begin{NBoutM}
\frac{1}{6}
\end{NBoutM}
Note that the equality between \code{Out[32]} and \code{Out[33]}
reflects the identity $f = F \circ X$, where $F$ is the chart function
(denoted \code{fU} above)
representing the scalar field $f$ on the chart $X$
(cf.\ Eq.~\eqref{e:man:repr_scalar_field}).

Internally, each chart function stores coordinate expressions with respect
to various computational backends:
\begin{itemize}
\item \Sage{} symbolic engine, based on the \soft{Pynac}\footnote{\url{http://pynac.org}}\index{Pynac} backend, with \soft{Maxima} used for some simplifications
or computation of integrals;
\item \soft{SymPy}\footnote{\url{https://www.sympy.org}}\index{SymPy} (Python library for symbolic mathematics);
\item in the future, more symbolic or numerical backends will be implemented.
\end{itemize}
The coordinate expressions are stored in the private dictionary \code{\_express}\footnote{not to be confused with
the attribute \code{\_express} of class \code{DiffScalarField} presented
at \code{In~[26]}}
of the class \code{ChartFunction},
whose keys are strings identifying the computational backends. By default
only \Sage{} symbolic expressions, i.e.\ expressions pertaining
to the so-called \Sage{}'s Symbolic Ring (\code{SR}),
are stored:
\begin{NBin}
fU._express
\end{NBin}
\begin{NBoutM}
\left\{\verb|SR| : \frac{1}{x^{2} + y^{2} + 1}\right\}
\end{NBoutM}
The public access to the private dictionary \code{\_express} is performed via the
method \code{expr()}:
\begin{NBin}
fU.expr()
\end{NBin}
\begin{NBoutM}
\frac{1}{x^{2} + y^{2} + 1}
\end{NBoutM}
\begin{NBin}
type(fU.expr())
\end{NBin}
\begin{NBout}
\begin{verbatim}
<type 'sage.symbolic.expression.Expression'>
\end{verbatim}
\end{NBout}
Actually, \code{fU.expr()} is a shortcut for \code{fU.expr('SR')} since
\code{SR} is the default symbolic backend. Note that the class
\code{Expression} is that devoted to \Sage{} symbolic expressions.
The method \code{expr()} can also be invoked to get the expression in
another symbolic backend, for instance \soft{SymPy}:
\begin{NBin}
fU.expr('sympy')
\end{NBin}
\begin{NBout}
\verb|1/(x**2|\phantom{\verb!x!}\verb|+|\phantom{\verb!x!}\verb|y**2|\phantom{\verb!x!}\verb|+|\phantom{\verb!x!}\verb|1)|
\end{NBout}
\begin{NBin}
type(fU.expr('sympy'))
\end{NBin}
\begin{NBout}
\begin{verbatim}
<class 'sympy.core.power.Pow'>
\end{verbatim}
\end{NBout}
This operation has updated the internal dictionary \code{\_express}
(compare with \code{Out~[34]}):
\begin{NBin}
fU._express
\end{NBin}
\begin{NBoutM}
\left\{\texttt{SR} : \frac{1}{x^{2} + y^{2} + 1},
\texttt{sympy}: \mbox{\texttt{1/(x**2 + y**2 + 1)}}\right\}
\end{NBoutM}
The default calculus backend for chart functions of chart \code{XU} can
changed thanks to the method \code{set\_calculus\_method()}:
\begin{NBin}
XU.set_calculus_method('sympy')
fU.expr()
\end{NBin}
\begin{NBout}
\verb|1/(x**2|\phantom{\verb!x!}\verb|+|\phantom{\verb!x!}\verb|y**2|\phantom{\verb!x!}\verb|+|\phantom{\verb!x!}\verb|1)|
\end{NBout}
Reverting to \Sage{}'s symbolic engine:
\begin{NBin}
XU.set_calculus_method('SR')
fU.expr()
\end{NBin}
\begin{NBoutM}
\frac{1}{x^{2} + y^{2} + 1}
\end{NBoutM}
Symbolic expressions can be accessed directly from the scalar field,
\code{f.expr(XU)} being a shortcut for \code{f.coord\_function(XU).expr()}:
\begin{NBin}
f.expr(XU)
\end{NBin}
\begin{NBoutM}
\frac{1}{x^{2} + y^{2} + 1}
\end{NBoutM}
\begin{NBin}
f.expr(XV)
\end{NBin}
\begin{NBoutM}
\frac{{x'}^{2} + {y'}^{2}}{{x'}^{2} + {y'}^{2} + 1}
\end{NBoutM}

\begin{figure}
\begin{center}
\begin{tikzpicture}[font=\small, node distance=0.5cm, minimum
height=2em, auto]

\node[native](unique_representation)
{\code{UniqueRepresentation}};

\node[native, right=of unique_representation](parent)
{\code{Parent}};

\coordinate (Middle) at ($(unique_representation)!0.5!(parent)$);

\node[diff, below=1.5cm of Middle](scalar_field_algebra)
{\code{ScalarFieldAlgebra}\\
{\scriptsize {\it element:} \code{ScalarField}}};
\path[line] (scalar_field_algebra) -- (unique_representation);
\path[line] (scalar_field_algebra) -- node [near start, yshift=-0.5em,
xshift=15.4em]
{\scriptsize {\it category:} \code{CommutativeAlgebras(base\_field)}} (parent);

\node[diff, below=1.25cm of scalar_field_algebra](diff_scalar_field_algebra)
{\code{DiffScalarFieldAlgebra}\\
{\scriptsize {\it element:} \code{DiffScalarField}}};
\path[line] (diff_scalar_field_algebra) -- (scalar_field_algebra);


\node[native, right=3cm of parent](caelement)
{\code{CommutativeAlgebraElement}};

\node[diff, below=1.25cm of caelement](scalarfield)
{\code{ScalarField}\\ \scriptsize{\it parent:} \code{ScalarFieldAlgebra}};
\path[line] (scalarfield) -- (caelement);

\node[diff, below=1.25cm of scalarfield](diff_scalarfield)
{\code{DiffScalarField}\\ \scriptsize{\it parent:} \code{DiffScalarFieldAlgebra}};
\path[line] (diff_scalarfield) -- (scalarfield);


\node[native_legend, below=0.5cm of diff_scalar_field_algebra]
(native_legend){};
\node[empty, right=0.5em of native_legend]
{Generic \Sage{} class};

\node[diff_legend, below=1.em of native_legend]
(diff_legend){};
\node[empty, right=0.5em of diff_legend]
{\SM{} class\\ \footnotesize (differential part)};

\end{tikzpicture}
\end{center}
\caption{\label{f:man:scalar_classes}\footnotesize
\Sage{} classes for scalar fields on a manifold.}
\end{figure}
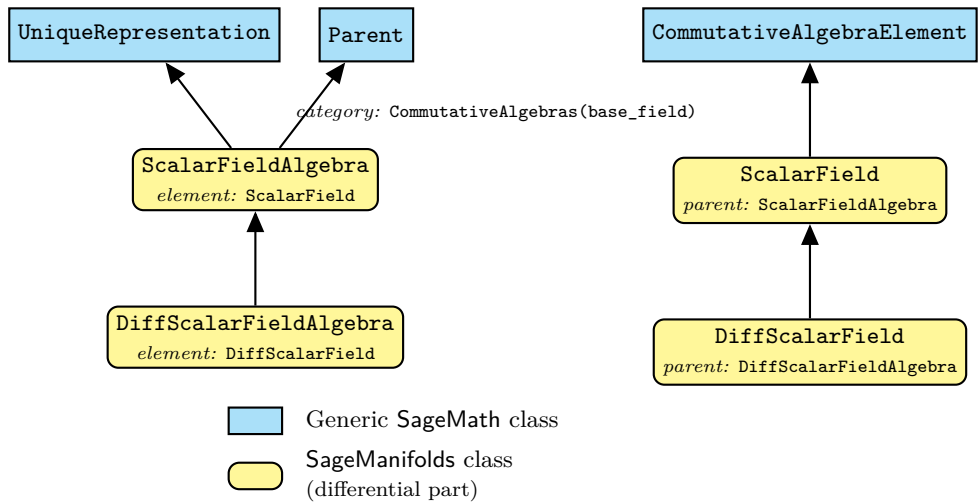

\subsection{Scalar field algebra} \label{s:man:scal_algebra}

The set $C^\infty(M)$
of all scalar fields on $M$ has naturally the structure of a
commutative algebra over $\K$: it is clearly a vector
space over $\K$ and it is endowed with a commutative ring structure
by pointwise multiplication:
\be
\forall f, g \in C^\infty(M),\quad \forall p\in M,\quad
(f.g)(p) := f(p) g(p) .
\ee
The algebra $C^\infty(M)$ is implemented in \Sage{} via the parent
class\\ \code{DiffScalarFieldAlgebra},\footurl{http://doc.sagemath.org/html/en/reference/manifolds/sage/manifolds/differentiable/scalarfield_algebra.html} in the category
\code{CommutativeAlgebras}. The corresponding element class
is of course \code{DiffScalarField} (cf.\ Fig.~\ref{f:man:scalar_classes}).

The \Sage{} object representing $C^\infty(M)$ is obtained from \code{M} via the
method\\ \code{scalar\_field\_algebra()}:
\begin{NBin}
CM = M.scalar_field_algebra()
CM
\end{NBin}
\begin{NBoutM}
C^{\infty}\left(M\right)
\end{NBoutM}
\begin{NBin}
CM.category()
\end{NBin}
\begin{NBoutM}
\mathbf{CommutativeAlgebras}_{\text{SR}}
\end{NBoutM}
As for the manifold classes, the actual Python class implementing
$C^\infty(M)$ is inherited from \code{DiffScalarFieldAlgebra} via \Sage{}'s
category framework (cf.\ Sec.~\ref{s:bas:def_manif}), hence it bares the name \code{DiffScalarFieldAlgebra\_with\_category}:
\begin{NBin}
type(CM)
\end{NBin}
\begin{NBout}
\begin{verbatim}
<class 'sage.manifolds.differentiable.scalarfield_algebra.
        DiffScalarFieldAlgebra_with_category'>
\end{verbatim}
\end{NBout}
The class \code{DiffScalarFieldAlgebra\_with\_category} is dynamically generated
as a subclass of \code{DiffScalarFieldAlgebra} with extra functionalities, like
for instance the method \code{is\_commutative()}:
\begin{NBin}
CM.is_commutative()
\end{NBin}
\begin{NBout}
\texttt{True}
\end{NBout}
To have a look at the corresponding code,
we use the double question mark, owing to the fact that \Sage{} is open-source:
\begin{NBin}
CM.is_commutative??
\end{NBin}
\begin{lstlisting}
def is_commutative(self):
    """
    Return ``True``, since commutative magmas are commutative.

    EXAMPLES::

        sage: Parent(QQ,category=CommutativeRings()).is_commutative()
        True
    """
    return True
File: .../local/lib/python2.7/site-packages/sage/categories/magmas.py
\end{lstlisting}
We see from the \code{File} field in line~11 that the code belongs to the category part of \Sage{}, not to the
manifold part, where the class \code{DiffScalarFieldAlgebra} is defined.
This shows that the method \code{is\_commutative()} has indeed be
added to the methods of the base class \code{DiffScalarFieldAlgebra}, while
dynamically generating the class \\ \code{DiffScalarFieldAlgebra-with-category}.

Regarding the scalar field \code{f} introduced in Sec.~\ref{s:man:def_scalar}, we
have of course
\begin{NBin}
f in CM
\end{NBin}
\begin{NBout}
\texttt{True}
\end{NBout}
Actually, in \Sage{} language, \code{CM}=$C^\infty(M)$ is the parent of \code{f}:
\begin{NBin}
f.parent() is CM
\end{NBin}
\begin{NBout}
\texttt{True}
\end{NBout}
The zero element of the algebra $C^\infty(M)$ is
\begin{NBin}
CM.zero().display()
\end{NBin}
\begin{NBoutM}
\begin{array}{llcl} 0:& M & \longrightarrow & \mathbb{R} \\ \mbox{on}\ U : & \left(x, y\right) & \longmapsto & 0 \\ \mbox{on}\ V : & \left({x'}, {y'}\right) & \longmapsto & 0 \end{array}
\end{NBoutM}
while its unit element is
\begin{NBin}
CM.one().display()
\end{NBin}
\begin{NBoutM}
\begin{array}{llcl} 1:& M & \longrightarrow & \mathbb{R} \\ \mbox{on}\ U : & \left(x, y\right) & \longmapsto & 1 \\ \mbox{on}\ V : & \left({x'}, {y'}\right) & \longmapsto & 1 \end{array}
\end{NBoutM}

\subsection{Implementation of algebra operations} \label{s:man:add_implement}

Let us consider some operation in the algebra $C^\infty(M)$:
\begin{NBin}
h = f + 2*CM.one()
h.display()
\end{NBin}
\begin{NBoutM}
\begin{array}{llcl} & M & \longrightarrow & \mathbb{R} \\ \mbox{on}\ U : & \left(x, y\right) & \longmapsto & \frac{2 \, x^{2} + 2 \, y^{2} + 3}{x^{2} + y^{2} + 1} \\ \mbox{on}\ V : & \left({x'}, {y'}\right) & \longmapsto & \frac{3 \, {x'}^{2} + 3 \, {y'}^{2} + 2}{{x'}^{2} + {y'}^{2} + 1} \end{array}
\end{NBoutM}
\begin{NBin}
h(p)
\end{NBin}
\begin{NBoutM}
\frac{13}{6}
\end{NBoutM}
Let us examine how the addition in \code{In~[53]} is performed. For the Python interpreter
\code{h = f + 2*CM.one()} is equivalent to \code{h = f.\_\_add\_\_(2*CM.one())},
i.e.\ the \code{+} operator amounts to calling the method \code{\_\_add\_\_()} on its
left operand, with the right operand as argument.
To have a look at the source code of this method, we use the double question mark:\footnote{In this
transcript of code and in those that follow, some parts
have been skipped, being not relevant for the discussion; they are marked
by ``\code{...}''.}
\begin{NBin}
f.__add__??
\end{NBin}
\label{p:man:list___add__}
\begin{lstlisting}
File: .../src/sage/structure/element.pyx
def __add__(left, right):
    """
    Top-level addition operator for :class:`Element` invoking
    the coercion model.

    See :ref:`element_arithmetic`.
    ...
    """
    cdef int cl = classify_elements(left, right)
    if HAVE_SAME_PARENT(cl):
        return (<Element>left)._add_(right)
    # Left and right are Sage elements => use coercion model
    if BOTH_ARE_ELEMENT(cl):
        return coercion_model.bin_op(left, right, add)
    ...
\end{lstlisting}
From lines 1 and 4, we
see that the method \code{\_\_add\_\_()} is implemented at the level
of the class \code{Element} from which \code{DiffScalarField} inherits, via
\code{CommutativeAlgebraElement} (cf.\ Fig.~\ref{f:man:scalar_classes}).
In the present case, \code{left} = \code{f} and \code{right} = \code{2*CM.one()}
have the same parent, namely the algebra \code{CM}, so that the actual
result is computed in line~12. The latter invokes the method \code{\_add\_()}
(note the single underscore on each side of \code{add}). This operator is
implemented at the level of \code{ScalarField}, as checked from the source code (see line~24 below):
\begin{NBin}
f._add_??
\end{NBin}
\begin{lstlisting}
def _add_(self, other):
    """
    Scalar field addition.

    INPUT:
    - ``other`` -- a scalar field (in the same algebra as ``self``)

    OUTPUT:
    - the scalar field resulting from the addition of ``self`` and
      ``other``
    ...
    """
    ...
    # Generic case:
    com_charts = self.common_charts(other)
    if com_charts is None:
        raise ValueError("no common chart for the addition")
    result = type(self)(self.parent())
    for chart in com_charts:
        # ChartFunction addition:
        result._express[chart] = self._express[chart] + other._express[chart]
    ...
    return result
File: .../local/lib/python2.7/site-packages/sage/manifolds/scalarfield.py
\end{lstlisting}
This reflects a general strategy\footnote{See \url{http://doc.sagemath.org/html/en/thematic_tutorials/coercion_and_categories.html} for details.} in \Sage{}: the arithmetic Python operators
\code{\_\_add\_\_()}, \code{\_\_sub\_\_()}, etc. are implemented at the
top-level class \code{Element}, while specific element subclasses,
like \code{ScalarField} here, implement single-underscore methods
\code{\_add\_()}, \code{\_sub\_()}, etc., which perform the actual computation
when both operands have the same parent.
Looking at the code (lines~15 to 23), we notice that the first step is to search
for the charts in which both operands of the addition operator have a coordinate
expression (line~15). This is performed by the method \code{common\_charts()};
in the current example, we get the two stereographic charts defined on $M$:
\begin{NBin}
f.common_charts(2*CM.one())
\end{NBin}
\begin{NBoutM}
\left[\left(U,(x, y)\right), \left(V,({x'}, {y'})\right)\right]
\end{NBoutM}
In general, \code{common\_charts()} returns the charts for which both operands
have already a known coordinate expression or for which a coordinate
expression can be computed by a known transition map, as we can see on the source
code:
\begin{NBin}
f.common_charts??
\end{NBin}
\begin{lstlisting}
def common_charts(self, other):
    """
    Find common charts for the expressions of the scalar field and
    ``other``.

    INPUT:
    - ``other`` -- a scalar field

    OUTPUT:
    - list of common charts; if no common chart is found, ``None`` is
      returned (instead of an empty list)
    ...
    """
    if not isinstance(other, ScalarField):
        raise TypeError("the second argument must be a scalar field")
    coord_changes = self._manifold._coord_changes
    resu = []
    #
    # 1/ Search for common charts among the existing expressions, i.e.
    #    without performing any expression transformation.
    #    -------------------------------------------------------------
    for chart1 in self._express:
        if chart1 in other._express:
            resu.append(chart1)
    # Search for a subchart:
    known_expr1 = self._express.copy()
    known_expr2 = other._express.copy()
    for chart1 in known_expr1:
        if chart1 not in resu:
            for chart2 in known_expr2:
                if chart2 not in resu:
                    if chart2 in chart1._subcharts:
                        self.expr(chart2)
                        resu.append(chart2)
                    if chart1 in chart2._subcharts:
                        other.expr(chart1)
                        resu.append(chart1)
    #
    # 2/ Search for common charts via one expression transformation
    #    ----------------------------------------------------------
    for chart1 in known_expr1:
        if chart1 not in resu:
            for chart2 in known_expr2:
                if chart2 not in resu:
                    if (chart1, chart2) in coord_changes:
                        self.coord_function(chart2, from_chart=chart1)
                        resu.append(chart2)
                    if (chart2, chart1) in coord_changes:
                        other.coord_function(chart1, from_chart=chart2)
                        resu.append(chart1)
    if resu == []:
        return None
    else:
        return resu
File: .../local/lib/python2.7/site-packages/sage/manifolds/scalarfield.py
\end{lstlisting}
Once the list of charts in which both operands have a coordinate expression
has been found,
the addition is performed at the chart function level (cf.\ Sec.~\ref{s:man:def_scalar}),
via the loop on the charts in lines~19-21 of the code for \code{\_add\_()}.
The code for the addition of chart functions defined on the same chart
is (recall that \code{fU} is the chart function representing $f$ in chart \code{XU}):
\begin{NBin}
fU._add_??
\end{NBin}
\begin{lstlisting}
def _add_(self, other):
    """
    Addition operator.

    INPUT:
    - ``other`` -- a :class:`ChartFunction` or a value

    OUTPUT:
    - chart function resulting from the addition of ``self``
      and ``other``
    ...
    """
    curr = self._calc_method._current
    res = self._simplify(self.expr() + other.expr())
    if curr =='SR' and res.is_trivial_zero():
        # NB: "if res == 0" would be too expensive (cf. #22859)
        return self.parent().zero()
    else:
        return type(self)(self.parent(), res)
File: .../local/lib/python2.7/site-packages/sage/manifolds/chart_func.py
\end{lstlisting}
We notice that the addition is performed in line~14 on the symbolic expression
with respect to the symbolic backend currently at work (\Sage{}/\soft{Pynac}, \soft{SymPy}, ...), as returned by
the method \code{expr()} (see Sec.~\ref{s:man:def_scalar}).
Let us recall that the user can change the symbolic backend at any time
by means of the method \code{set\_calculus\_method()}, applied either to
a chart or to an open subset (possibly \code{M} itself).
Besides, we notice on line~14 above that the result of the symbolic addition
is automatically simplified, by means of the method \code{\_simplify}.
The latter invokes a chain of simplifying functions, which depends on the
symbolic backend.\footnote{See
\url{https://github.com/sagemath/sage/blob/develop/src/sage/manifolds/utilities.py}
for details; note that the simplifications regarding the \soft{SymPy} engine are not
fully implemented yet.}

Let us now discuss the second case in the \code{\_\_add\_\_()} method of
\code{Element}, namely the case for which the parents of both operands are
different (lines~14-15 in the code listed as a result of \code{In [55]},
on page~\pageref{p:man:list___add__}). This case is treated via \Sage{} coercion\index{coercion} model, which allows one to deal with additions like
\begin{NBin}
h1 = f + 2
h1.display()
\end{NBin}
\begin{NBoutM}
\begin{array}{llcl} & M & \longrightarrow & \mathbb{R} \\ \mbox{on}\ U : & \left(x, y\right) & \longmapsto & \frac{2 \, x^{2} + 2 \, y^{2} + 3}{x^{2} + y^{2} + 1} \\ \mbox{on}\ V : & \left({x'}, {y'}\right) & \longmapsto & \frac{3 \, {x'}^{2} + 3 \, {y'}^{2} + 2}{{x'}^{2} + {y'}^{2} + 1} \end{array}
\end{NBoutM}
A priori, \code{f + 2} is not a well defined operation, since the integer $2$ does not
belong to the algebra $C^\infty(M)$. However \Sage{} manages to treat it
because $2$ can be coerced (i.e.\ automatically and unambiguously converted) via \code{CM(2)}
into a element of $C^\infty(M)$, namely the constant scalar field whose value is $2$:
\begin{NBin}
CM(2).display()
\end{NBin}
\begin{NBoutM}
\begin{array}{llcl} & M & \longrightarrow & \mathbb{R} \\ \mbox{on}\ U : & \left(x, y\right) & \longmapsto & 2 \\ \mbox{on}\ V : & \left({x'}, {y'}\right) & \longmapsto & 2 \end{array}
\end{NBoutM}
This happens because there exists a coercion map from the parent of $2$, namely the ring of integers $\mathbb{Z}$
(denoted \code{ZZ} in \Sage{}), to $C^\infty(M)$:
\begin{NBin}
2.parent()
\end{NBin}
\begin{NBoutM}
\mathbf{Z}
\end{NBoutM}
\begin{NBin}
CM.has_coerce_map_from(ZZ)
\end{NBin}
\begin{NBout}
\texttt{True}
\end{NBout}

\chapter{Vector fields} \label{s:vec}

\minitoc

\section{Introduction}

This chapter is devoted to the most basic objects of tensor calculus:
vector fields. We start by defining tangent vectors and tangent spaces
on a differentiable manifold (Sec.~\ref{s:vec:tangent_vectors}), and then
move to vector fields (Sec.~\ref{s:vec:vector_fields}).

\section{Tangent vectors} \label{s:vec:tangent_vectors}

\subsection{Definitions} \label{s:vec:def_tangent_vector}

Let $\M$ be a smooth manifold of dimension $n$ over the topological field $\K$
and $C^\infty(M)$ the corresponding algebra of scalar fields introduced in Sec.~\ref{s:man:scal_algebra}.
For $p\in M$, a \defin{tangent vector at}\index{tangent!vector} $p$ is
a map
\be
    \w{v}: C^\infty(M) \longrightarrow \K
\ee
such that (i) $\w{v}$ is $\K$-linear and (ii) $\w{v}$ obeys
\be \label{e:vec:derivation}
    \forall f,g \in C^\infty(M),\quad
        \w{v}(fg) = \w{v}(f) g(p) + f(p) \w{v}(g) .
\ee
Because of property~\eqref{e:vec:derivation}, one says that $\w{v}$ is
a \defin{derivation at} $p$.

The set $T_p\M$ of all tangent vectors at $p$ is a vector space of dimension
$n$ over $\K$; it is called the \defin{tangent space to} $\M$
\defin{at}\index{tangent!space} $p$.

\subsection{SageMath implementation} \label{s:vec:tangent_impl}

To illustrate the implementation of tangent vectors in \Sage{}, we shall
consider the same example $M=\Sp$ as in Chap.~\ref{s:man}. First of all,
we recreate the same objects as in Chap.~\ref{s:man}, starting with the manifold
$M$ and its two stereographic charts $X_U = (U,(x,y))$ and $X_V = (V,(x',y'))$,
with $M=U\cup V$ (the full Jupyter notebook is available at
\url{https://sagemanifolds.obspm.fr/jncf2018/}):
\setcounter{NBin}{0}
\begin{NBin}
%display latex
\end{NBin}
\begin{NBin}
M = Manifold(2, 'M')
U = M.open_subset('U')
XU.<x,y> = U.chart()
V = M.open_subset('V')
XV.<xp,yp> = V.chart("xp:x' yp:y'")
M.declare_union(U,V)
XU_to_XV = XU.transition_map(XV,
                             (x/(x^2+y^2), y/(x^2+y^2)),
                             intersection_name='W',
                             restrictions1= x^2+y^2!=0,
                             restrictions2= xp^2+yp^2!=0)
XV_to_XU = XU_to_XV.inverse()
M.atlas()
\end{NBin}
\begin{NBoutM}
\left[\left(U,(x, y)\right), \left(V,({x'}, {y'})\right), \left(W,(x, y)\right), \left(W,({x'}, {y'})\right)\right]
\end{NBoutM}
Then we introduce the point $p\in U$ of coordinates $(x,y)=(1,2)$:
\begin{NBin}
p = U((1,2), chart=XU, name='p')
print(p)
\end{NBin}
\begin{NBprint}
Point p on the 2-dimensional differentiable manifold M
\end{NBprint}
The canonical embedding of $\Sp$ in $\R^3$ is defined mostly for
graphical purposes:
\begin{NBin}
R3 = Manifold(3, 'R^3', r'\mathbb{R}^3')
XR3.<X,Y,Z> = R3.chart()
Phi = M.diff_map(R3, {(XU, XR3):
                       [2*x/(1+x^2+y^2), 2*y/(1+x^2+y^2),
                       (x^2+y^2-1)/(1+x^2+y^2)],
                      (XV, XR3):
                       [2*xp/(1+xp^2+yp^2), 2*yp/(1+xp^2+yp^2),
                       (1-xp^2-yp^2)/(1+xp^2+yp^2)]},
                  name='Phi', latex_name=r'\Phi')
Phi.display()
\end{NBin}
\begin{NBoutM}
\begin{array}{llcl} \Phi:& M & \longrightarrow & \mathbb{R}^3 \\ \mbox{on}\ U : & \left(x, y\right) & \longmapsto & \left(X, Y, Z\right) = \left(\frac{2 \, x}{x^{2} + y^{2} + 1}, \frac{2 \, y}{x^{2} + y^{2} + 1}, \frac{x^{2} + y^{2} - 1}{x^{2} + y^{2} + 1}\right) \\ \mbox{on}\ V : & \left({x'}, {y'}\right) & \longmapsto & \left(X, Y, Z\right) = \left(\frac{2 \, {x'}}{{x'}^{2} + {y'}^{2} + 1}, \frac{2 \, {y'}}{{x'}^{2} + {y'}^{2} + 1}, -\frac{{x'}^{2} + {y'}^{2} - 1}{{x'}^{2} + {y'}^{2} + 1}\right) \end{array}
\end{NBoutM}
\begin{NBin}
graph = XU.plot(chart=XR3, mapping=Phi, number_values=25,
                label_axes=False) + \
        XV.plot(chart=XR3, mapping=Phi, number_values=25,
                color='green', label_axes=False) + \
        p.plot(chart=XR3, mapping=Phi, label_offset=0.05)
show(graph, viewer='threejs', online=True)
\end{NBin}
\begin{center}
\includegraphics[width=0.8\textwidth]{sphere_stereo.png}
\end{center}
Finally, the last objects defined in Chap.~\ref{s:man} are the
scalar field $f$:
\begin{NBin}
f = M.scalar_field({XU: 1/(1+x^2+y^2), XV: (xp^2+yp^2)/(1+xp^2+yp^2)},
                   name='f')
f.display()
\end{NBin}
\begin{NBoutM}
\begin{array}{llcl} f:& M & \longrightarrow & \mathbb{R} \\ \mbox{on}\ U : & \left(x, y\right) & \longmapsto & \frac{1}{x^{2} + y^{2} + 1} \\ \mbox{on}\ V : & \left({x'}, {y'}\right) & \longmapsto & \frac{{x'}^{2} + {y'}^{2}}{{x'}^{2} + {y'}^{2} + 1} \end{array}
\end{NBoutM}
and its parent, namely the commutative algebra $C^\infty(\M)$ of
smooth maps $\M\to\R$:
\begin{NBin}
CM = M.scalar_field_algebra()
CM
\end{NBin}
\begin{NBoutM}
C^{\infty}\left(M\right)
\end{NBoutM}

The tangent space at the point $p$ introduced in \code{In~[3]} is generated by
\begin{NBin}
Tp = M.tangent_space(p)
Tp
\end{NBin}
\begin{NBoutM}
T_{p}\,M
\end{NBoutM}
It is a vector space over $\K$ (here $\K=\R$, which is represented by \Sage{}'s Symbolic
Ring \code{SR}):
\begin{NBin}
print(Tp.category())
\end{NBin}
\begin{NBprint}
Category of finite dimensional vector spaces over Symbolic Ring
\end{NBprint}
The dimension of the vector space $T_p\M$ equals that of the manifold $\M$:
\begin{NBin}
dim(Tp)
\end{NBin}
\begin{NBoutM}
2
\end{NBoutM}
Tangent spaces are implemented as a class inherited from \code{TangentSpace}
via the category framework:
\begin{NBin}
type(Tp)
\end{NBin}
\begin{NBout}
\begin{verbatim}
<class
 'sage.manifolds.differentiable.tangent_space.TangentSpace_with_category'>
\end{verbatim}
\end{NBout}
\code{FiniteRankFreeModule},\footurl{http://doc.sagemath.org/html/en/reference/tensor_free_modules/sage/tensor/modules/finite_rank_free_module.html}, which, in \Sage{} is devoted to free modules of finite rank
without any distinguished basis:
\begin{NBin}
isinstance(Tp, FiniteRankFreeModule)
\end{NBin}
\begin{NBout}
\texttt{True}
\end{NBout}
\begin{remark}
In \Sage{}, free modules with a distinguished basis
are created with the command \code{FreeModule} or \code{VectorSpace}
and belong to classes different from \code{FiniteRankFreeModule}.
The differences are illustrated at\\
{\scriptsize \url{http://doc.sagemath.org/html/en/reference/modules/sage/tensor/modules/finite_rank_free_module.html#diff-freemodule}}.
\end{remark}

Two bases of $T_p\M$ are already available: those generated by the derivations
at $p$ along the coordinates of charts \code{XU} and \code{XV} respectively:
\begin{NBin}
Tp.bases()
\end{NBin}
\begin{NBoutM}
\left[\left(\frac{\partial}{\partial x },\frac{\partial}{\partial y }\right), \left(\frac{\partial}{\partial {x'} },\frac{\partial}{\partial {y'} }\right)\right]
\end{NBoutM}
None of these bases is distinguished, but one if the default one, which
simply means that it is the basis to be considered if the basis argument
is skipped in some methods:
\begin{NBin}
Tp.default_basis()
\end{NBin}
\begin{NBoutM}
\left(\frac{\partial}{\partial x },\frac{\partial}{\partial y }\right)
\end{NBoutM}

A tangent vector is created as an element of the tangent space by the
standard \Sage{} procedure\index{parent}\index{element}
\code{new\_element = parent(...)}, where \code{...}
stands for some material sufficient to construct the element:
\begin{NBin}
vp = Tp((-3, 2), name='v')
print(vp)
\end{NBin}
\begin{NBprint}
Tangent vector v at Point p on the 2-dimensional differentiable manifold M
\end{NBprint}
Since the basis is not specified, the pair $(-3,2)$ refers to components
with respect to the default basis:
\begin{NBin}
vp.display()
\end{NBin}
\begin{NBoutM}
v = -3 \frac{\partial}{\partial x } + 2 \frac{\partial}{\partial y }
\end{NBoutM}
We have of course
\begin{NBin}
vp.parent()
\end{NBin}
\begin{NBoutM}
T_{p}\,M
\end{NBoutM}
\begin{NBin}
vp in Tp
\end{NBin}
\begin{NBout}
\texttt{True}
\end{NBout}
As other manifold objects, tangent vectors have some plotting capabilities:
\begin{NBin}
graph += vp.plot(chart=XR3, mapping=Phi, scale=0.5, color='gold')
show(graph, viewer='threejs', online=True)
\end{NBin}
\begin{center}
\includegraphics[width=0.8\textwidth]{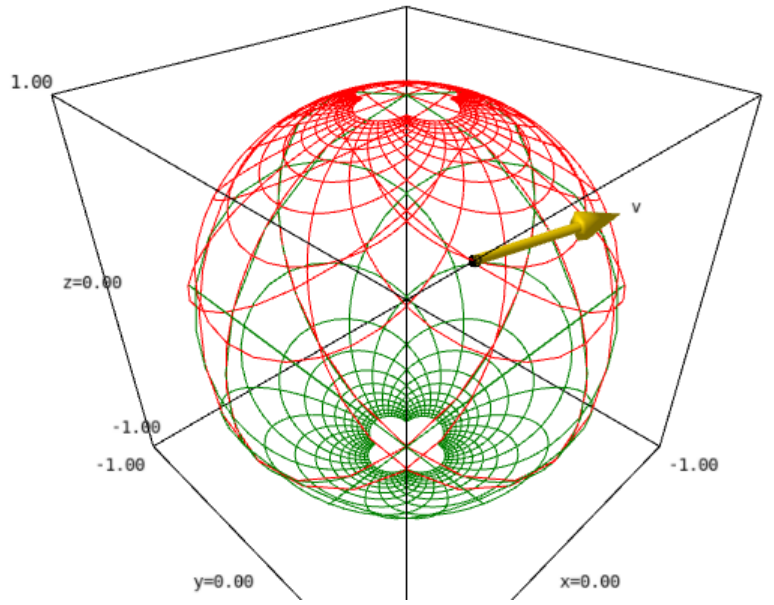}
\end{center}

The main attribute of the object \code{vp} representing the vector $\w{v}$ is the
private dictionary \code{\_components}, which stores the components of $\w{v}$ in various bases of $T_p M$:
\begin{NBin}
vp._components
\end{NBin}
\begin{NBout}
$\displaystyle
\bigg\{\left(\frac{\partial}{\partial x },\frac{\partial}{\partial y }\right) :$\\
\texttt{1-index components w.r.t.~Basis (d/dx,d/dy) on the Tangent space at Point p on the 2-dimensional differentiable manifold M}
$\bigg\}$
\end{NBout}
The keys of the dictionary \code{\_components} are the bases of $T_p M$, while the values belong to the class \code{Components}\footurl{http://doc.sagemath.org/html/en/reference/tensor_free_modules/sage/tensor/modules/comp.html} devoted to store ring elements indexed by integers or tuples of integers:
\begin{NBin}
vpc = vp._components[Tp.default_basis()]
vpc
\end{NBin}
\begin{NBout}
\texttt{1-index components w.r.t.~Basis (d/dx,d/dy) on the Tangent space at Point p
on the 2-dimensional differentiable manifold M}
\end{NBout}
\begin{NBin}
type(vpc)
\end{NBin}
\begin{NBout}
\begin{verbatim}
<class 'sage.tensor.modules.comp.Components'>
\end{verbatim}
\end{NBout}
The components themselves are stored in the private dictionary \code{\_comp} of the \code{Components} object, with the indices as keys:
\begin{NBin}
vpc._comp
\end{NBin}
\begin{NBoutM}
\left\{\left(0\right) : -3, \left(1\right) : 2\right\}
\end{NBoutM}
Hence the components are not stored via a sequence data type (list or tuple), as
one might have expected, but via a mapping type (dictionary). This is a general
feature of the class \code{Components} and all its subclasses, which permits
to not store vanishing components and, in case of symmetries (for multi-index
objects like tensors), to store only non-redundant components.


\section{Vector fields} \label{s:vec:vector_fields}

\subsection{Definition}

The \defin{tangent bundle}\index{tangent!bundle} of $\M$ is the disjoint union
of the tangent spaces at all points of $\M$:
\be
    T\M = \coprod_{p\in\M} T_p \M .
\ee
Elements of $T\M$ are usually denoted by $(p, \w{u})$, with $\w{u}\in T_p\M$.
The tangent bundle is canonically endowed with the
\defin{projection map}\index{projection!map}:
\be
    \begin{array}{cccc}
    \pi : & T\M & \longrightarrow & \M \\
        & (p,\w{u}) & \longmapsto & p .
    \end{array}
\ee

The tangent bundle inherits some manifold structure from $\M$:
$T\M$ is a smooth manifold of dimension $2n$ over $\K$ ($n$ being the dimension
of $\M$).

A \defin{vector field}\index{vector!field} on $\M$ is a continuous
right-inverse of the projection map, i.e.\ it is a map
\be
    \begin{array}{cccc}
    \w{v} : & \M & \longrightarrow & T\M \\
        & p & \longmapsto & \left. \w{v}\right| _p
    \end{array}
\ee
such that $\pi\circ \w{v} = \mathrm{Id}_{\M}$, i.e.\ such that
\be
    \forall p\in \M, \quad \left. \w{v}\right| _p \in T_p\M .
\ee

\subsection{Module of vector fields} \label{s:vec:vector_module}

The set $\X(\M)$ of all vector fields on $\M$
is naturally endowed with two algebraic structures:
\begin{enumerate}
\item $\X(\M)$ is a (infinite dimensional) vector space over $\K$ --- the base field of $\M$ ---,
the scalar multiplication $\K\times\X(M)\to \X(M)$, $(\lambda,\w{v})\mapsto \lambda \w{v}$
being defined by
\be
    \forall p\in\M,
    \quad \left. (\lambda \w{v}) \right| _p =  \left. \lambda \w{v} \right| _p,
\ee
where
the right-hand side involves the scalar multiplication
in the vector space $T_p\M$;
\item $\X(\M)$ is a module\index{module} over $C^\infty(M)$ --- the commutative
algebra of scalar fields ---,
the scalar multiplication $C^\infty(M)\times\X(M)\to \X(M)$, $(f,\w{v})\mapsto f \w{v}$
being defined by
\be
    \forall p\in\M,
    \quad \left. (f \w{v}) \right| _p = \left.f(p)  \w{v} \right| _p ,
\ee
where
the right-hand side involves the scalar multiplication by $f(p)\in \K$
in the vector space $T_p\M$.
\end{enumerate}
An important subcase of 2 is when $\X(\M)$ is a
\defin{free module}\index{free module}\index{module!free --} over $C^\infty(M)$,
i.e.\ when $\X(\M)$ admits a basis\index{basis} (a generating set consisting of linearly independent elements). If this occurs, then $\X(\M)$ is actually a
\defin{free module of finite rank}\index{free module!of finite rank}\index{finite rank!free module of --} over $C^\infty(M)$ and its rank is $n$ -- the dimension of
$\M$ over $\K$, which means that all bases share the same cardinality, namely $n$.
One says that $\M$ is a \defin{parallelizable}\index{parallelizable} manifold.
A basis $(\w{e}_a)_{1\leq a \leq n}$ of $\X(\M)$ is called a
\defin{vector frame}\index{vector!frame};
for any $p\in\M$, $(\left.\w{e}_a \right| _p)_{1\leq a \leq n}$
is then a basis of the tangent vector space $T_p\M$.
Any vector field has a unique decomposition with respect to the vector
frame\footnote{Einstein's convention for summation on repeated indices is assumed.} $(\w{e}_a)_{1\leq a \leq n}$:
\be \label{e:vec:v_expand}
    \forall \w{v}\in\X(\M),\quad \w{v} = v^a \w{e}_a,\quad\mbox{with\ } v^a \in C^\infty(\M) .
\ee
At each point $p\in \M$, Eq.~\eqref{e:vec:v_expand} gives birth to an identity in
the tangent space $T_p \M$:
\be
    \left.\w{v} \right| _p = v^a(p)  \left.\w{e}_a \right| _p,\quad\mbox{with\ } v^a(p) \in \K ,
\ee
which is nothing but the expansion of the tangent vector $\left.\w{v} \right| _p$
on the basis $(\left.\w{e}_a \right| _p)_{1\leq a \leq n}$ of the
vector space $T_p \M$.

Note that if $\M$ is covered by a chart $X$, i.e.~$\M$ is the domain of
the chart $X$, then $\M$ is parallelizable and a vector frame
is $(\partial/\partial x^a)_{1\leq a \leq n}$, where the $x^a$'s are
the coordinates of chart $X$. Such a vector frame is called a
\defin{coordinate frame}\index{coordinate!frame} or \defin{natural basis}\index{natural basis}. More generally, examples of parallelizable manifolds are
\cite{Lee13}
\begin{itemize}
\item the Cartesian space $\mathbb{R}^n$ for $n=1,2,\ldots$,
\item the circle $\mathbb{S}^1$,
\item the torus $\mathbb{T}^2 = \mathbb{S}^1\times \mathbb{S}^1$,
\item the sphere $\mathbb{S}^3 \simeq \mathrm{SU}(2)$, as any Lie group,
\item the sphere $\mathbb{S}^7$,
\item any orientable 3-manifold (Steenrod theorem~\cite{Steen51}).
\end{itemize}
On the other hand, examples of non-parallelizable manifolds are
\begin{itemize}
\item the sphere $\mathbb{S}^2$ (as a consequence of the hairy ball theorem),
as well as any sphere $\mathbb{S}^n$ with $n\not\in\{1,3,7\}$,
\item the real projective plane $\mathbb{RP}^2$.
\end{itemize}
Actually, ``most'' manifolds are non-parallelizable.
As noticed above, if a manifold is covered by a single chart, it is
parallelizable (the prototype being $\mathbb{R}^n$). But the reverse is not
true: $\mathbb{S}^1$ and $\mathbb{T}^2$ are parallelizable and require
at least two charts to cover them.

\subsection{SageMath implementation} \label{s:vec:vector_field_impl}

Among the two algebraic structures for $\X(\M)$ discussed in Sec.~\ref{s:vec:vector_module},
we select the second one, i.e.\ we consider $\X(\M)$ as a $C^\infty(M)$-module.
With respect to the infinite-dimensional $\K$-vector space point of view,
the advantage for the implementation is the reduction to finite-dimensional structures: free modules of rank $n$ on parallelizable open subsets of $\M$.
Indeed, if $U$ is such an open subset, i.e.\ if $\X(U)$ is a free $C^\infty(U)$-module
of rank $n$, the generic class \code{FiniteRankFreeModule} discussed in
Sec.~\ref{s:vec:tangent_impl} can be used to implement $\X(U)$. The great benefit
is that all calculus implemented on the free module elements, like the addition
or the scalar multiplication, can be used as such for vector fields.
This implies that vector fields will be described by their (scalar-field) components
on vector frames, as defined by Eq.~\eqref{e:vec:v_expand}, on parallelizable
open subsets of $\M$.

If the manifold $\M$ is not parallelizable,
we assume that it can be covered by a finite number $m$
of parallelizable open subsets $U_i$ ($1\leq i \leq m$):
\be
    M = \bigcup_{i=1}^m U_i, \qquad\mbox{with}\quad U_i\ \mbox{parallelizable}
\ee
In particular, this holds if $\M$ is compact, for any compact
manifold admits a finite atlas.

\begin{figure}
\begin{center}
\begin{tikzpicture}[font=\footnotesize, node distance=0.75cm, minimum
height=2em, auto]

\node[native](unique_representation)
{\code{UniqueRepresentation}};

\node[native, right=of unique_representation](parent)
{\code{Parent}};

\coordinate (Middle) at ($(unique_representation)!0.5!(parent)$);

\node[diff, below=3cm of Middle](vectorfieldmodule)
{\code{VectorFieldModule}\\
{\scriptsize {\it ring:} \code{DiffScalarFieldAlgebra}}\\
{\scriptsize {\it element:} \code{VectorField}}};

\path[line] (vectorfieldmodule) -- (unique_representation);
\path[line] (vectorfieldmodule) -- node [near start, yshift=4.5em,
xshift=2.8em, rotate=65]
{\footnotesize {\it category:} \code{Modules}} (parent);

\node[diff, left=of vectorfieldmodule](tensorfieldmodule)
{\code{TensorFieldModule}\\
{\scriptsize {\it ring:} \code{DiffScalarFieldAlgebra}}\\
{\scriptsize {\it element:} \code{TensorField}}};

\path[line] (tensorfieldmodule) -- (unique_representation);
\path[line] (tensorfieldmodule) -- node [near start, xshift=5em,
yshift=2.2em, rotate=29]
{\footnotesize {\it category:} \code{Modules}} (parent);

\node[diff, below=of vectorfieldmodule](vectorfieldfreemodule)
{\code{VectorFieldFreeModule}\\
{\scriptsize {\it ring:} \code{DiffScalarFieldAlgebra}}\\
{\scriptsize {\it element:} \code{VectorFieldParal}}};

\node[diff, below=of vectorfieldfreemodule](tensorfieldfreemodule)
{\code{TensorFieldFreeModule}\\
{\scriptsize {\it ring:} \code{DiffScalarFieldAlgebra}}\\
{\scriptsize {\it element:} \code{TensorFieldParal}}};

\node[alg, right=1cm of vectorfieldmodule](finiterankfreemodule)
{\code{FiniteRankFreeModule}\\
{\scriptsize {\it ring:} \code{CommutativeRing}}\\
{\scriptsize {\it element:} \code{FiniteRankFreeModuleElement}}};

\node[alg, right=of vectorfieldfreemodule](tensorfreemodule)
{\code{TensorFreeModule}\\
{\scriptsize {\it element:}}\\
{\scriptsize \code{FreeModuleTensor}}};

\node[diff, right=of tensorfreemodule](tangentspace)
{\code{TangentSpace}\\
{\scriptsize {\it ring:} \code{SR}}\\
{\scriptsize {\it element:}}\\
{\scriptsize \code{TangentVector}}};

\path[line] (finiterankfreemodule) -- (unique_representation);
\path[line] (finiterankfreemodule) -- node [near start,
xshift=2.2em, rotate=-43]
{\footnotesize {\it category:} \code{Modules}} (parent);

\path[line] (tangentspace) -- (finiterankfreemodule);
\path[line] (tensorfreemodule) -- (finiterankfreemodule);

\path[line] (vectorfieldfreemodule) -- (finiterankfreemodule);
\path[line] (tensorfieldfreemodule) -- (tensorfreemodule);

\node[native_legend, left=4cm of vectorfieldfreemodule]
(native_legend){};
\node[empty, right=0.5em of native_legend]
{Generic \Sage{} class};

\node[alg_legend, below=1.em of native_legend]
(alg_legend){};
\node[empty, right=0.5em of alg_legend]
{\SM{} class\\ \footnotesize (algebraic part)};

\node[diff_legend, below=1.5em of alg_legend]
(diff_legend){};
\node[empty, right=0.5em of diff_legend]
{\SM{} class\\ \footnotesize (differential part)};

\end{tikzpicture}
\end{center}
\caption{\label{f:vec:module_classes} \footnotesize
\Sage{} classes for modules involved in differentiable manifolds.}
\end{figure}
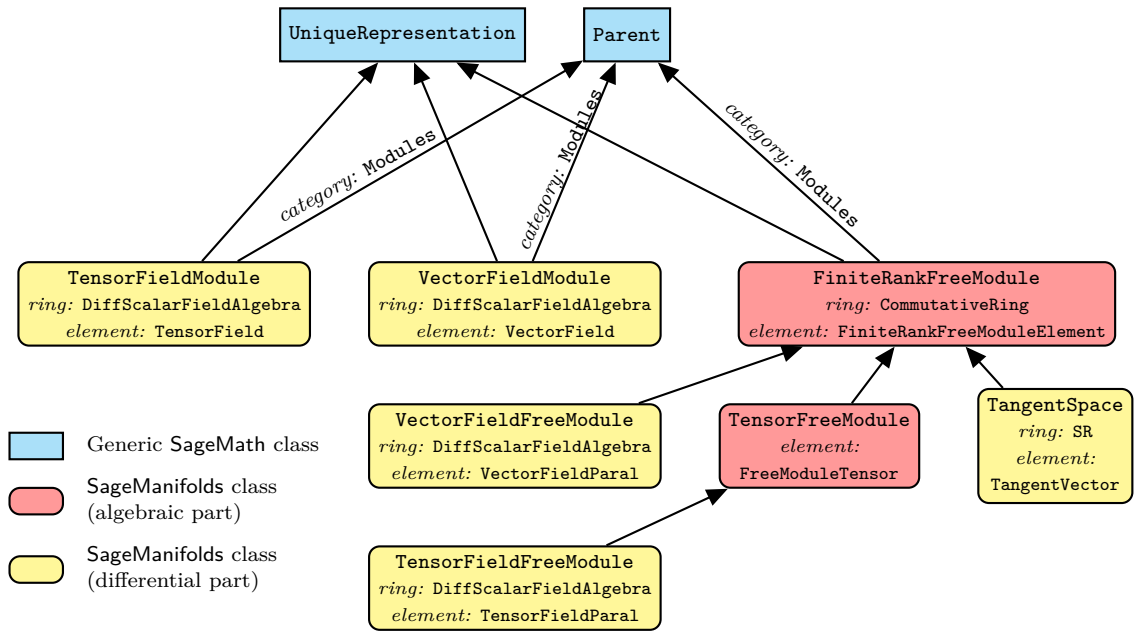

For each $i\in\{1,\ldots,m\}$, $\X(U_i)$ is a free module of rank $n=\dim  M$ and is implemented in SageMath as an instance of
\code{VectorFieldFreeModule}, which is a subclass of
\code{FiniteRankFreeModule}. This inheritance is illustrated in
Fig.~\ref{f:vec:module_classes}. On that figure, we note that the class
\code{TangentSpace} discussed in Sec.~\ref{s:vec:tangent_impl} inherits from\\
\code{FiniteRankFreeModule} as well.

A vector field $\w{v}\in\X(M)$ is then described by its
restrictions $\left(\left. \w{v}\right| _{U_i}\right)_{1\leq i \leq m}$ to each of the $U_i$'s.
Assuming that at least one vector frame is introduced in each of the $U_i$'s,
$(\w{e}_{i,a})_{1\leq a \leq n}$ say, the restriction $\left. \w{v}\right| _{U_i}$ of
$\w{v}$ to $U_i$ is decribed by its components $v_i^a$ in that frame:
\be \label{e:vec:vi_expand}
    \left. \w{v}\right| _{U_i} = v_i^a \, \w{e}_{i,a},\quad\mbox{with\ } v_i^a \in C^\infty(U_i) .
\ee

Let us illustrate this strategy with the example of $\Sp$.
We get $\X(\M)$ by\footnote{We are using \code{YM} to denote $\X(\M)$ and not \code{XM}, because we reserve the symbol \code{X} to denote coordinate charts, as
\code{XU}, \code{XV} or \code{XR3}.}
\begin{NBin}
YM = M.vector_field_module()
YM
\end{NBin}
\begin{NBoutM}
\mathfrak{X}\left(M\right)
\end{NBoutM}
As discussed above, $\X(\M)$ is considered as a module over $C^\infty(M)$:
\begin{NBin}
YM.category()
\end{NBin}
\begin{NBoutM}
\mathbf{Modules}_{C^{\infty}\left(M\right)}
\end{NBoutM}
Since the algebra $C^\infty(M)$ is denoted \code{CM}, we have
\begin{NBin}
YM.base_ring() is CM
\end{NBin}
\begin{NBout}
\texttt{True}
\end{NBout}
$\X(\M)$ is not a free module; in particular, we can check that its \Sage{} implementation does not
belong to the class \code{FiniteRankFreeModule}:
\begin{NBin}
isinstance(YM, FiniteRankFreeModule)
\end{NBin}
\begin{NBout}
\texttt{False}
\end{NBout}
This is because $M=\Sp$ is not a parallelizable manifold:
\begin{NBin}
M.is_manifestly_parallelizable()
\end{NBin}
\begin{NBout}
\texttt{False}
\end{NBout}
Via \Sage{} category framework,
the module $\X(\M)$ is implemented by a dynamically-generated subclass
of the class \code{VectorFieldModule}, which is devoted to modules of vector fields
on non-parallelizable manifolds:
\begin{NBin}
type(YM)
\end{NBin}
\begin{NBout}
\begin{verbatim}
<class 'sage.manifolds.differentiable.vectorfield_module.
        VectorFieldModule_with_category'>
\end{verbatim}
\end{NBout}
On the contrary, the set $\mathfrak{X}(U)$ of vector fields on $U$ is a free module of
finite rank over the algebra $C^\infty(U)$:
\begin{NBin}
YU = U.vector_field_module()
isinstance(YU, FiniteRankFreeModule)
\end{NBin}
\begin{NBout}
\texttt{True}
\end{NBout}
\begin{NBin}
YU.base_ring()
\end{NBin}
\begin{NBoutM}
C^{\infty}\left(U\right)
\end{NBoutM}
This is because the open subset $U$ is a parallelizable manifold:
\begin{NBin}
U.is_manifestly_parallelizable()
\end{NBin}
\begin{NBout}
\texttt{True}
\end{NBout}
being the domain of a coordinate chart:
\begin{NBin}
U.is_manifestly_coordinate_domain()
\end{NBin}
\begin{NBout}
\texttt{True}
\end{NBout}
We can check that in $U$'s atlas, at least one chart has $U$ for domain:
\begin{NBin}
U.atlas()
\end{NBin}
\begin{NBoutM}
\left[\left(U,(x, y)\right), \left(W,(x, y)\right), \left(W,({x'}, {y'})\right)\right]
\end{NBoutM}
This chart is \code{XU} = $(U, (x,y))$, i.e.\ the chart of stereographic coordinates
from the North pole.
The rank of $\X(U)$ as a free $C^\infty(U)$-module is the manifold's dimension:
\begin{NBin}
rank(YU)
\end{NBin}
\begin{NBoutM}
2
\end{NBoutM}
Via the category framework,
the free module $\X(U)$ is implemented by a dynamically-generated subclass
of the class \code{VectorFieldFreeModule}, which is devoted to modules of vector fields
on parallelizable manifolds:
\begin{NBin}
type(YU)
\end{NBin}
\begin{NBout}
\begin{verbatim}
<class 'sage.manifolds.differentiable.vectorfield_module.
        VectorFieldFreeModule_with_category'>
\end{verbatim}
\end{NBout}
The class \code{VectorFieldFreeModule} is itself a subclass
of the generic class\\ \code{FiniteRankFreeModule}:
\begin{NBin}
class_graph(
  sage.manifolds.differentiable.vectorfield_module.VectorFieldFreeModule
).plot()
\end{NBin}
\begin{center}
\includegraphics[width=0.7\textwidth]{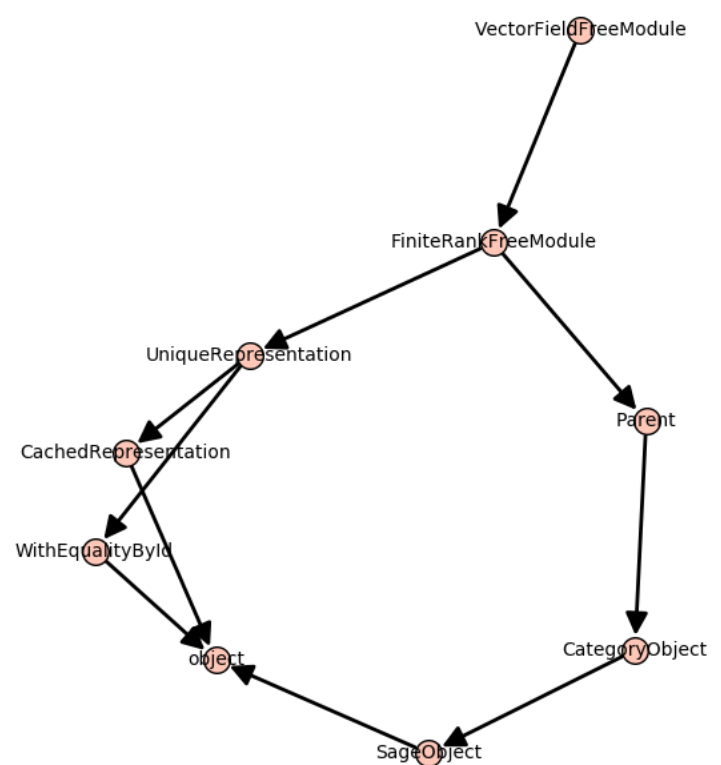}
\end{center}

Since $U$ is a chart domain, the free module $\X(U)$ is automatically endowed with a basis,
which is the coordinate frame associated to the chart:
\begin{NBin}
YU.bases()
\end{NBin}
\begin{NBoutM}
\left[\left(U, \left(\frac{\partial}{\partial x },\frac{\partial}{\partial y }\right)\right)\right]
\end{NBoutM}
Let us denote by \code{eU} this frame. We can set \code{eU = YU.bases()[0]} or
alternatively
\begin{NBin}
eU = YU.default_basis()
eU
\end{NBin}
\begin{NBoutM}
\left(U, \left(\frac{\partial}{\partial x },\frac{\partial}{\partial y }\right)\right)
\end{NBoutM}
Another equivalent instruction would have been \code{eU = U.default\_frame()}.

Similarly, $\X(V)$ is a free module, endowed with the coordinate frame
associated to stereographic coordinates from the South pole, which we
denote by \code{eV}:
\begin{NBin}
YV = V.vector_field_module()
YV.bases()
\end{NBin}
\begin{NBoutM}
\left[\left(V, \left(\frac{\partial}{\partial {x'} },\frac{\partial}{\partial {y'} }\right)\right)\right]
\end{NBoutM}
\begin{NBin}
eV = YV.default_basis()
eV
\end{NBin}
\begin{NBoutM}
\left(V, \left(\frac{\partial}{\partial {x'} },\frac{\partial}{\partial {y'} }\right)\right)
\end{NBoutM}

If we consider the intersection $W=U\cap V$, we notice its module
of vector fields is endowed with two bases, reflecting the fact that
$W$ is covered by two charts: $(W,(x,y))$ and $(W,(x',y'))$:
\begin{NBin}
W = U.intersection(V)
YW = W.vector_field_module()
YW.bases()
\end{NBin}
\begin{NBoutM}
\left[\left(W, \left(\frac{\partial}{\partial x },\frac{\partial}{\partial y }\right)\right), \left(W, \left(\frac{\partial}{\partial {x'} },\frac{\partial}{\partial {y'} }\right)\right)\right]
\end{NBoutM}
Let us denote by \code{eUW} and \code{eUV} these two bases, which are
actually the restrictions of the vector frames \code{eU} and \code{eV} to
$W$:
\begin{NBin}
eUW = eU.restrict(W)
eVW = eV.restrict(W)
YW.bases() == [eUW, eVW]
\end{NBin}
\begin{NBout}
\texttt{True}
\end{NBout}
The free module $\X(W)$ is also automatically endowed with automorphisms
connecting the two bases, i.e.\ change-of-frame operators:
\begin{NBin}
W.changes_of_frame()
\end{NBin}
\begin{NBout}
\[
\bigg\{\left(\left(W, \left(\frac{\partial}{\partial {x'} },\frac{\partial}{\partial {y'} }\right)\right), \left(W, \left(\frac{\partial}{\partial x },\frac{\partial}{\partial y }\right)\right)\right) :
\]
Field of tangent-space automorphisms on the Open subset W of the 2-dimensional differentiable manifold M,
\[
\left(\left(W, \left(\frac{\partial}{\partial x },\frac{\partial}{\partial y }\right)\right), \left(W, \left(\frac{\partial}{\partial {x'} },\frac{\partial}{\partial {y'} }\right)\right)\right) :
\]
Field of tangent-space automorphisms on the Open subset W of the 2-dimensional differentiable manifold M $\bigg\}$
\end{NBout}
The first of them is
\begin{NBin}
P = W.change_of_frame(eUW, eVW)
P
\end{NBin}
\begin{NBout}
\texttt{Field of tangent-space automorphisms on the Open subset W of the 2-dimensional
differentiable manifold M}
\end{NBout}
It belongs to the general linear group of the free module $\X(W)$:
\begin{NBin}
P.parent()
\end{NBin}
\begin{NBoutM}
\mathrm{GL}\left( \mathfrak{X}\left(W\right) \right)
\end{NBoutM}
and its matrix is deduced from the Jacobian matrix of the transition map
\code{XV} $\to$ \code{XU}:
\begin{NBin}
P[:]
\end{NBin}
\begin{NBoutM}
\left(\begin{array}{rr}
-x^{2} + y^{2} & -2 \, x y \\
-2 \, x y & x^{2} - y^{2}
\end{array}\right)
\end{NBoutM}

\subsection{Construction and manipulation of vector fields}

Let us introduce a vector field $\w{v}$ on $\M$:
\begin{NBin}
v = M.vector_field(name='v')
v[eU, 0] = f.restrict(U)
v[eU, 1] = -2
v.display(eU)
\end{NBin}
\begin{NBoutM}
v = \left( \frac{1}{x^{2} + y^{2} + 1} \right) \frac{\partial}{\partial x } -2 \frac{\partial}{\partial y }
\end{NBoutM}
Notice that, at this stage, we have defined $\w{v}$ only on $U$, by setting
its components in the vector frame \code{eU}, either explicitly as scalar
fields, like the component $v^0$ set to the restriction of $f$ to $U$ or
implicitly, like the component $v^1$: the integer \code{-2}
will be coerced to the constant scalar field of value $-2$ (cf.\ Sec.~\ref{s:man:add_implement}).
We can ask for the scalar-field value of a component via the double-bracket
operator; since \code{eU} is the default frame on $M$, we do not have to specify
it:
\begin{NBin}
v[[0]]
\end{NBin}
\begin{NBoutM}
f
\end{NBoutM}
\begin{NBin}
v[[0]].display()
\end{NBin}
\begin{NBoutM}
\begin{array}{llcl} f:& U & \longrightarrow & \mathbb{R} \\ & \left(x, y\right) & \longmapsto & \frac{1}{x^{2} + y^{2} + 1} \\ \mbox{on}\ W : & \left({x'}, {y'}\right) & \longmapsto & \frac{{x'}^{2} + {y'}^{2}}{{x'}^{2} + {y'}^{2} + 1} \end{array}
\end{NBoutM}
Note that, for convenience, the single bracket operator returns a chart function
of the component:
\begin{NBin}
v[0]
\end{NBin}
\begin{NBoutM}
\frac{1}{x^{2} + y^{2} + 1}
\end{NBoutM}
The restriction of $\w{v}$ to $W$ is of course
\begin{NBin}
v.restrict(W).display(eUW)
\end{NBin}
\begin{NBoutM}
v = \left( \frac{1}{x^{2} + y^{2} + 1} \right) \frac{\partial}{\partial x } -2 \frac{\partial}{\partial y }
\end{NBoutM}
Since we have a second vector frame on $W$, namely \code{eVW}, and the
change-of-frame automorphisms are known, we can ask for the components
of $\w{v}$ with respect to that frame:
\begin{NBin}
v.restrict(W).display(eVW)
\end{NBin}
\begin{NBout}
$\displaystyle
v = \left( \frac{4 \, x y^{3} - x^{2} + 4 \, {\left(x^{3} + x\right)} y + y^{2}}{x^{6} + y^{6} + {\left(3 \, x^{2} + 1\right)} y^{4} + x^{4} + {\left(3 \, x^{4} + 2 \, x^{2}\right)} y^{2}} \right) \frac{\partial}{\partial {x'} }$ \\
$\displaystyle
+ \left( -\frac{2 \, {\left(x^{4} - y^{4} + x^{2} + x y - y^{2}\right)}}{x^{6} + y^{6} + {\left(3 \, x^{2} + 1\right)} y^{4} + x^{4} + {\left(3 \, x^{4} + 2 \, x^{2}\right)} y^{2}} \right) \frac{\partial}{\partial {y'} }$
\end{NBout}
Notice that the components are expressed in terms of the coordinates $(x,y)$
since they form the default chart on $W$. To have them expressed in
terms of the coordinates $(x',y')$, we have to add the restriction of
the chart
$(V,(x',y'))$ to $W$ as the second argument of the method
\code{display()}:
\begin{NBin}
v.restrict(W).display(eVW, XV.restrict(W))
\end{NBin}
\begin{NBout}
$\displaystyle
v = \left( -\frac{{x'}^{4} - 4 \, {x'} {y'}^{3} - {y'}^{4} - 4 \, {\left({x'}^{3} + {x'}\right)} {y'}}{{x'}^{2} + {y'}^{2} + 1} \right) \frac{\partial}{\partial {x'} } $\\
$\displaystyle
+ \left( -\frac{2 \, {\left({x'}^{4} + {x'}^{3} {y'} + {x'} {y'}^{3} - {y'}^{4} + {x'}^{2} - {y'}^{2}\right)}}{{x'}^{2} + {y'}^{2} + 1} \right) \frac{\partial}{\partial {y'} }$
\end{NBout}
We extend the expression of $\w{v}$ to the full vector frame \code{XV}
by continuation of this expression:
\begin{NBin}
v.add_comp_by_continuation(eV, W, chart=XV)
\end{NBin}
\ \\
We have then
\begin{NBin}
v.display(eV)
\end{NBin}
\begin{NBout}
$\displaystyle
v = \left( -\frac{{x'}^{4} - 4 \, {x'} {y'}^{3} - {y'}^{4} - 4 \, {\left({x'}^{3} + {x'}\right)} {y'}}{{x'}^{2} + {y'}^{2} + 1} \right) \frac{\partial}{\partial {x'} }$\\
$\displaystyle
 + \left( -\frac{2 \, {\left({x'}^{4} + {x'}^{3} {y'} + {x'} {y'}^{3} - {y'}^{4} + {x'}^{2} - {y'}^{2}\right)}}{{x'}^{2} + {y'}^{2} + 1} \right) \frac{\partial}{\partial {y'} }$
\end{NBout}
At this stage, the vector field $\w{v}$ is defined in all $M$.
According to the hairy ball theorem\index{hairy ball theorem}, it has to vanish somewhere.
Let us show that this occurs at the North pole, by first introducing the
latter, as the point of stereographic coordinates $(x',y')=(0,0)$:
\begin{NBin}
N = M((0,0), chart=XV, name='N')
print(N)
\end{NBin}
\begin{NBprint}
Point N on the 2-dimensional differentiable manifold M
\end{NBprint}
As a check, we verify that the image of $N$ by the canonical embedding
$\Phi: \Sp \to \R^3$ is the point of Cartesian coordinates $(0,0,1)$:
\begin{NBin}
XR3(Phi(N))
\end{NBin}
\begin{NBoutM}
(0, 0, 1)
\end{NBoutM}
The vanishing of $\left.\w{v}\right| _N$:
\begin{NBin}
v.at(N).display()
\end{NBin}
\begin{NBoutM}
v = 0
\end{NBoutM}
On the other hand, $\w{v}$ does not vanish at the point $p$ introduced above:
\begin{NBin}
v.at(p).display()
\end{NBin}
\begin{NBoutM}
v = \frac{1}{6} \frac{\partial}{\partial x } -2 \frac{\partial}{\partial y }
\end{NBoutM}
We may plot the vector field $\w{v}$ in terms of the stereographic coordinates
from the North pole:
\begin{NBin}
v.plot(chart=XU, chart_domain=XU, max_range=2,
       number_values=5, scale=0.4, aspect_ratio=1)
\end{NBin}
\begin{center}
\includegraphics[width=0.6\textwidth]{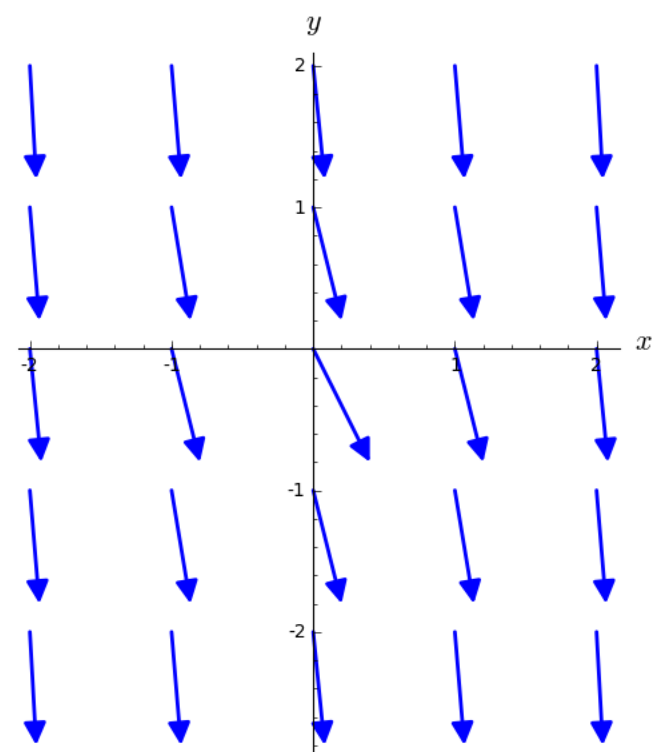}
\end{center}
or in term of those from the South pole:
\begin{NBin}
v.plot(chart=XV, chart_domain=XV, max_range=2,
       number_values=9, scale=0.05, aspect_ratio=1)
\end{NBin}
\begin{center}
\includegraphics[width=0.8\textwidth]{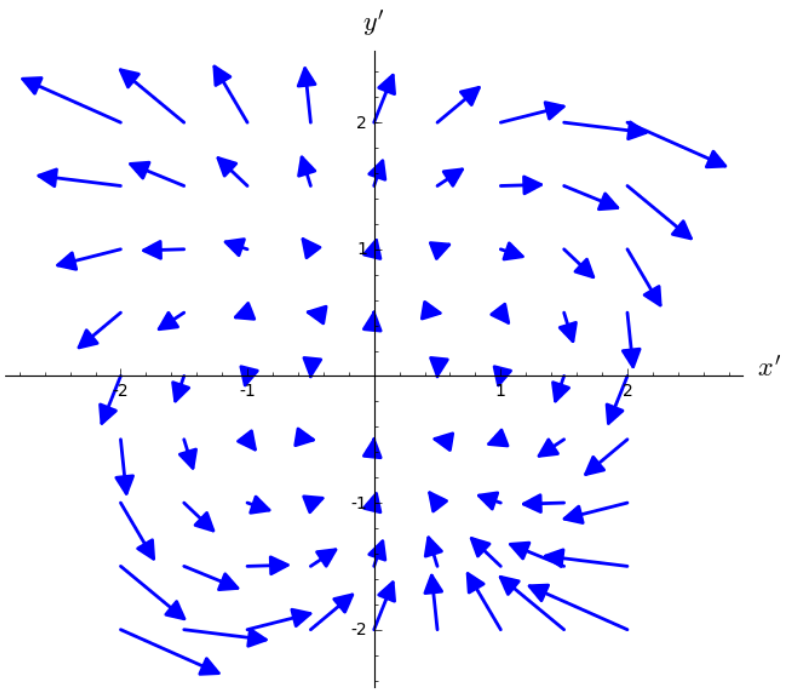}
\end{center}
Thanks to the embedding $\Phi$, we may also have a 3D plot of the vector
field $\w{v}$
atop of the 3D plot already obtained:
\begin{NBin}
graph_v = v.plot(chart=XR3, mapping=Phi, chart_domain=XU,
                 number_values=7, scale=0.2) + \
          v.plot(chart=XR3, mapping=Phi, chart_domain=XV,
                 number_values=7, scale=0.2)
show(graph + graph_v, viewer='threejs', online=True)
\end{NBin}
\begin{center}
\includegraphics[width=0.6\textwidth]{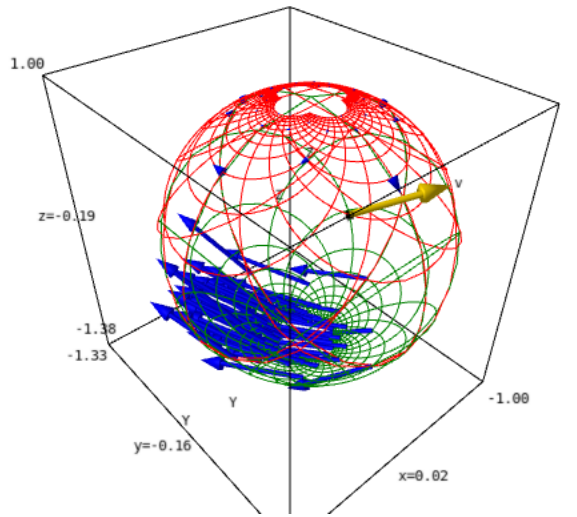}
\end{center}
Note that the sampling, performed on the two charts \code{XU} and \code{XV}
is not uniform on the sphere. A better sampling would be achieved by introducing
spherical coordinates.

\subsection{Implementation details regarding vector fields}

Let us now investigate some internals of the implementation of vector fields.
Vector fields on $M$ are implemented via the class
\code{VectorField}\footurl{http://doc.sagemath.org/html/en/reference/manifolds/sage/manifolds/differentiable/vectorfield.html} (actually by a dynamically generated subclass of it, within \Sage{} category
framework):
\begin{NBin}
isinstance(v, sage.manifolds.differentiable.vectorfield.VectorField)
\end{NBin}
\begin{NBout}
\texttt{True}
\end{NBout}

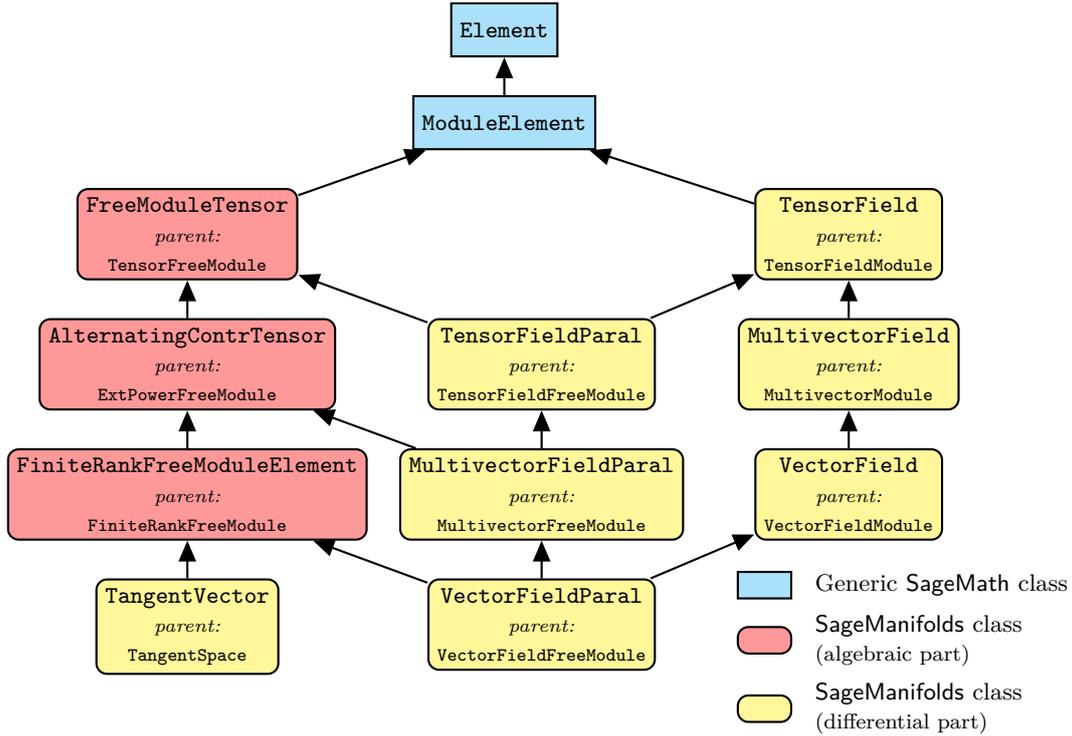
\begin{figure}
\begin{center}
\begin{tikzpicture}[font=\small, node distance=0.5cm, minimum
height=2em, auto]

\node[native](element)
{\code{Element}};

\node[native, below=of element](module_element)
{\code{ModuleElement}};
\path[line] (module_element) -- (element);

\node[alg, below left=0.5cm and 1.5cm of module_element](freemoduletensor)
{\code{FreeModuleTensor}\\ \scriptsize {\it parent:}\\ \scriptsize
\code{TensorFreeModule}};
\path[line] (freemoduletensor) -- (module_element);

\node[alg, below=of freemoduletensor](altcontrtensor)
{\code{AlternatingContrTensor}\\ \scriptsize {\it parent:}\\ \scriptsize
\code{ExtPowerFreeModule}};
\path[line] (altcontrtensor) -- (freemoduletensor);

\node[alg, below=of altcontrtensor](finiterankfreemoduleelement)
{\code{FiniteRankFreeModuleElement}\\ \scriptsize {\it parent:}\\ \scriptsize
\code{FiniteRankFreeModule}};
\path[line] (finiterankfreemoduleelement) -- (altcontrtensor);

\node[diff, below=of finiterankfreemoduleelement](tangentvector)
{\code{TangentVector}\\ \scriptsize {\it parent:}\\ \scriptsize
\code{TangentSpace}};
\path[line] (tangentvector) -- (finiterankfreemoduleelement);

\node[diff, right=6cm of freemoduletensor](tensorfield)
{\code{TensorField}\\ \scriptsize {\it parent:}\\ \scriptsize
\code{TensorFieldModule}};
\path[line] (tensorfield) -- (module_element);

\node[diff, below=of tensorfield](multivectorfield)
{\code{MultivectorField}\\ \scriptsize {\it parent:}\\ \scriptsize
\code{MultivectorModule}};
\path[line] (multivectorfield) -- (tensorfield);

\node[diff, below=of multivectorfield](vectorfield)
{\code{VectorField}\\ \scriptsize {\it parent:}\\ \scriptsize
\code{VectorFieldModule}};
\path[line] (vectorfield) -- (multivectorfield);

\node[diff, right=1.2cm of altcontrtensor](tensorfieldparal)
{\code{TensorFieldParal}\\ \scriptsize {\it parent:}\\ \scriptsize
\code{TensorFieldFreeModule}};
\path[line] (tensorfieldparal) -- (freemoduletensor);
\path[line] (tensorfieldparal) -- (tensorfield);

\node[diff, below=of tensorfieldparal](multivectorfieldparal)
{\code{MultivectorFieldParal}\\ \scriptsize {\it parent:}\\ \scriptsize
\code{MultivectorFreeModule}};
\path[line] (multivectorfieldparal) -- (altcontrtensor);
\path[line] (multivectorfieldparal) -- (tensorfieldparal);

\node[diff, below=of multivectorfieldparal](Vectorfieldparal)
{\code{VectorFieldParal}\\ \scriptsize {\it parent:}\\ \scriptsize
\code{VectorFieldFreeModule}};
\path[line] (Vectorfieldparal) -- (finiterankfreemoduleelement);
\path[line] (Vectorfieldparal) -- (multivectorfieldparal);
\path[line] (Vectorfieldparal) -- (vectorfield);

\node[native_legend, below left=0.4cm and -0.5cm of vectorfield]
(native_legend){};
\node[empty, right=0.5em of native_legend]
{Generic \Sage{} class};

\node[alg_legend, below=1.em of native_legend]
(alg_legend){};
\node[empty, right=0.5em of alg_legend]
{\SM{} class\\ \footnotesize (algebraic part)};

\node[diff_legend, below=1.5em of alg_legend]
(diff_legend){};
\node[empty, right=0.5em of diff_legend]
{\SM{} class\\ \footnotesize (differential part)};

\end{tikzpicture}
\end{center}
\caption{\label{f:vec:tensorfield_classes}\footnotesize
\Sage{} classes for tensor fields involved in differentiable manifolds.
There are various multiple inheritances involving diamond diagrams;
Python's method resolution order algorithm (MRO) relies on the ordering of the parents
in the class declaration and this order can be read from the left to the right in
this figure. For instance, the class \code{VectorFieldParal} is declared as
\code{class VectorFieldParal(FiniteRankFreeModuleElement, MultivectorFieldParal, VectorField)}.}
\end{figure}

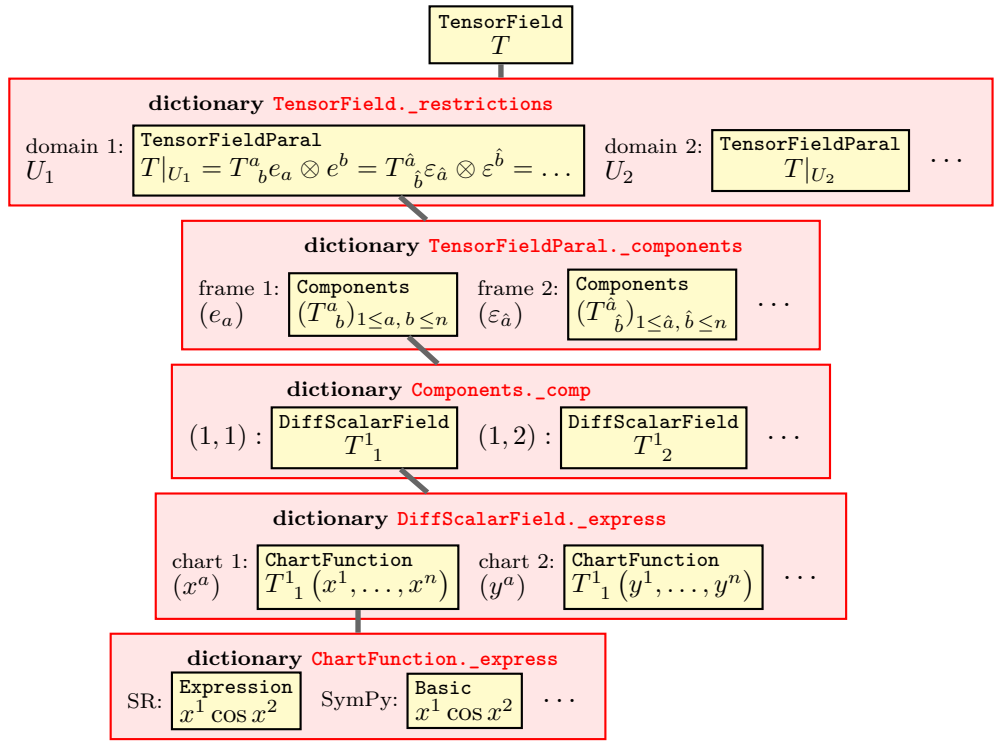
\begin{figure}
\begin{center}
\begin{tikzpicture}[font=\footnotesize, remember picture,
node distance=0.5em, minimum height=0.5em, auto]

\node[tens, align=center](tensorfield)
{\code{TensorField}\\ \normalsize $T$};

\node[dict, below=of tensorfield](restrictions){
\begin{tikzpicture}
  \node[empty, align=left](description)
  {\bf dictionary \textcolor{red}{\code{TensorField.\_restrictions}}};
  \node[empty, below left=of description](domain1)
  {domain 1:\\ {\normalsize $U_1$}};
  \node[tens, right=0em of domain1](tensorfield1)
  {\code{TensorFieldParal}\\
  {\normalsize $T|_{U_1}=T^a_{\ \, b} e_a\otimes
e^b=T^{\hat{a}}_{\ \, \hat{b}}\varepsilon_{\hat{a}}\otimes
\varepsilon^{\hat{b}}=\dots$}};
  \node[empty, right=of tensorfield1](domain2)
  {domain 2:\\ {\normalsize $U_2$}};
  \node[tens, align=center, right=0em of domain2](tensorfield2)
  {\code{TensorFieldParal}\\ {\normalsize $T|_{U_2}$}};
  \node[empty, right=of tensorfield2](more)
  {\large $\dots$};
\end{tikzpicture}
};

\node[dict, below=of restrictions](components){
\begin{tikzpicture}
  \node[empty, align=left](description)
  {\bf dictionary
\textcolor{red}{\code{TensorFieldParal.\_components}}};
  \node[empty, below left=of description](frame1)
  {frame 1:\\ {\normalsize $(e_a)$}};
  \node[tens, right=0em of frame1](components1)
  {\code{Components}\\
  {\normalsize $(T^a_{\ \, b})_{1\le a,\,b\,\le n}$}};
  \node[empty, right=of components1](frame2)
  {frame 2:\\ {\normalsize $(\varepsilon_{\hat{a}})$}};
  \node[tens, right=0em of frame2](components2)
  {\code{Components}\\
  {\normalsize $(T^{\hat{a}}_{\ \, \hat{b}})_{1\le
\hat{a},\,\hat{b}\,\le n}$}};
  \node[empty, right=of components2](more)
  {\large $\dots$};
\end{tikzpicture}
};

\node[dict, below=of components](comp){
\begin{tikzpicture}
  \node[empty, align=left](description)
  {\bf dictionary \textcolor{red}{\code{Components.\_comp}}};
  \node[empty, below left=of description](comp1)
  {\normalsize $(1,1):$};
  \node[tens, align=center, right=0em of comp1](scalarfield1)
  {\code{DiffScalarField}\\
  {\normalsize $T^1_{\ \, 1}$}};
  \node[empty, right=of scalarfield1](comp2)
  {\normalsize $(1,2):$};
  \node[tens, align=center, right=0em of comp2](scalarfield2)
  {\code{DiffScalarField}\\
  {\normalsize $T^1_{\ \, 2}$}};
  \node[empty, right=of scalarfield2](more)
  {\large $\dots$};
\end{tikzpicture}
};

\node[dict, below=of comp](express){
\begin{tikzpicture}
  \node[empty, align=left](description)
  {\bf dictionary \textcolor{red}{\code{DiffScalarField.\_express}}};
  \node[empty, below left=of description](chart1)
  {chart 1:\\
  {\normalsize $\left(x^a\right)$}};
  \node[tens, right=0em of chart1](functionchart1)
  {\code{ChartFunction}\\
  {\normalsize $T^1_{\ \, 1}\left(x^1,\dots,x^n\right)$}};
  \node[empty, right=of functionchart1](chart2)
  {chart 2:\\
  {\normalsize $\left(y^a\right)$}};
  \node[tens, right=0em of chart2](functionchart2)
  {\code{ChartFunction}\\
  {\normalsize $T^1_{\ \, 1}\left(y^1,\dots,y^n\right)$}};
  \node[empty, right=of functionchart2](more)
  {\large $\dots$};
\end{tikzpicture}
};

\node[dict, below=0.3cm of functionchart1](chartfexpress){
\begin{tikzpicture}
  \node[empty, align=left](description)
  {\bf dictionary \textcolor{red}{\code{ChartFunction.\_express}}};
  \node[empty, below left=of description](methSR)
  {SR:};
  \node[tens, right=0em of methSR](exprSR)
  {\code{Expression}\\
  {\normalsize $x^1\cos x^2$}};
  \node[empty, right=of exprSR](methSymPy)
  {SymPy:};
  \node[tens, right=0em of methSymPy](exprSymPy)
  {\code{Basic}\\
  {\normalsize $x^1\cos x^2$}};
  \node[empty, right=of exprSymPy](more)
  {\large $\dots$};
\end{tikzpicture}
};

\draw[thick, line width=0.2em, black!60,-] (restrictions) -- (tensorfield);
\draw[thick, line width=0.2em, black!60,-] (components) -- (tensorfield1);
\draw[thick, line width=0.2em, black!60,-] (comp) -- (components1);
\draw[thick, line width=0.2em, black!60,-] (express) -- (scalarfield1);
\draw[thick, line width=0.2em, black!60,-] (chartfexpress) -- (functionchart1);

\end{tikzpicture}
\end{center}
\caption{\label{f:vec:storage_tensor} \footnotesize
Internal storage of tensor fields. Red boxes
represent Python dictionaries, yellow boxes are dictionary values, with the corresponding
dictionary key located on the left of them.
The Python class of each dictionary value is indicated in typewriter font at the top of the
yellow box. In the hierarchical tree, only the leftmost
branch is indicated by grey connectors. In the special case of vector fields, the classes
\code{TensorField} and \code{TensorFieldParal} are to be replaced by \code{VectorField} and \code{VectorFieldParal} respectively.}
\end{figure}

Since $M$ is not parallelizable, the defining data of a
vector field $\w{v}$ on $M$ are its restrictions
$\left(\left. \w{v}\right| _{U_i}\right)_{1\leq i \leq m}$
to parallelizable open subsets $U_i$,
following the scheme presented in Sec.~\ref{s:vec:vector_field_impl}.
These restrictions are stored in the private dictionary \code{\_restrictions}, whose keys are
the open subsets:
\begin{NBin}
v._restrictions
\end{NBin}
\begin{NBoutM}
\left\{V : v, W : v, U : v\right\}
\end{NBoutM}
Let us consider one of these restrictions, for instance the restriction
$\left. \w{v}\right| _U$ to $U$:
\begin{NBin}
vU = v._restrictions[U]
vU is v.restrict(U)
\end{NBin}
\begin{NBout}
\texttt{True}
\end{NBout}
Since $U$ is a parallelizable open subset, the object \code{vU} belongs
to the class \code{VectorFieldParal}, which is devoted to vector fields
on parallelizable manifolds:
\begin{NBin}
isinstance(vU, sage.manifolds.differentiable.vectorfield.VectorFieldParal)
\end{NBin}
\begin{NBout}
\texttt{True}
\end{NBout}
The class \code{VectorFieldParal} inherits both from
\code{FiniteRankFreeModuleElement} (as \code{TangentVector}) and from
\code{VectorField} (see Fig.~\ref{f:vec:tensorfield_classes}).
The defining data of $\left. \w{v}\right| _U$ are
its sets of components with respect to (possibly various)
vector frames on $U$, according to Eq.~\eqref{e:vec:vi_expand}. The sets of components are stored in the private dictionary \code{\_components}, whose keys are the vector frames:
\begin{NBin}
vU._components
\end{NBin}
\begin{NBoutM}
\left\{\left(U, \left(\frac{\partial}{\partial x },\frac{\partial}{\partial y }\right)\right) :
\mbox{\texttt{1-index components w.r.t.~Coordinate frame (U, (d/dx,d/dy))}}\right\}
\end{NBoutM}
Similarly, we have:
\begin{NBin}
v._restrictions[W]._components
\end{NBin}
\begin{NBout}
\[
\bigg\{\left(W, \left(\frac{\partial}{\partial x },\frac{\partial}{\partial y }\right)\right) :
\mbox{\texttt{1-index components w.r.t.~Coordinate frame (W, (d/dx,d/dy))}},
\]
\[
\left(W, \left(\frac{\partial}{\partial {x'} },\frac{\partial}{\partial {y'} }\right)\right) :
\mbox{\texttt{1-index components w.r.t.~Coordinate frame (W, (d/dxp,d/dyp)}}
\bigg\}
\]
\end{NBout}
The values of the dictionary \code{\_components} belong to the same class
\code{Components} as that discussed in Sec.~\ref{s:vec:tangent_impl} for
the storage of components of tangent vectors:
\begin{NBin}
vUc = vU._components[eU]
vUc
\end{NBin}
\begin{NBout}
\texttt{1-index components w.r.t.~Coordinate frame (U, (d/dx,d/dy))}
\end{NBout}
\begin{NBin}
type(vUc)
\end{NBin}
\begin{NBout}
\begin{verbatim}
<class 'sage.tensor.modules.comp.Components'>
\end{verbatim}
\end{NBout}
As already mentioned in Sec.~\ref{s:vec:tangent_impl}, the components themselves are stored
in the private attribute \code{\_comp} of the \code{Components} object; this is a dictionary
whose keys are the indices:
\begin{NBin}
vUc._comp
\end{NBin}
\begin{NBout}
$\displaystyle
\big\{\left(0\right) : f, $\\
$\displaystyle
\phantom{x}\left(1\right) : \mbox{Scalar field on the Open subset U of the 2-dimensional differentiable manifold M}\big\}$
\end{NBout}
The difference with the tangent vector case is that the values of that dictionary are now scalar fields, i.e.\ elements of $C^\infty(U)$ in the present case. This is of course in agreement with the treatment of $\mathfrak{X}(U)$ as a free module over $C^\infty(U)$,
as discussed in Sec.~\ref{s:vec:vector_field_impl}.
Taking into account the storage of scalar fields presented in Sec.~\ref{s:man:def_scalar},
the full storage structure of vector fields is presented in Fig.~\ref{f:vec:storage_tensor}
(the latter actually regards tensor fields, of which vector fields constitute a subcase).

Let us perform some algebraic operation on vector fields:
\begin{NBin}
w = v + f*v
w
\end{NBin}
\begin{NBout}
Vector field on the 2-dimensional differentiable manifold M
\end{NBout}
The code for the addition is accessible via
\begin{NBin}
v.__add__??
\end{NBin}
\begin{lstlisting}
File: .../src/sage/structure/element.pyx
def __add__(left, right):
    """
    Top-level addition operator for :class:`Element` invoking
    the coercion model.

    See :ref:`element_arithmetic`.
    ...
    """
    cdef int cl = classify_elements(left, right)
    if HAVE_SAME_PARENT(cl):
        return (<Element>left)._add_(right)
    # Left and right are Sage elements => use coercion model
    if BOTH_ARE_ELEMENT(cl):
        return coercion_model.bin_op(left, right, add)
    ...
\end{lstlisting}
This is exactly the same method \code{\_\_add\_\_()} as that discussed in
Sec.~\ref{s:man:add_implement} for the addition of scalar fields (cf.\ page~\pageref{p:man:list___add__}), namely
the method \code{\_\_add\_\_()} of the top-level class \code{Element}, from
which both \code{VectorField} and \code{DiffScalarField} inherit, cf.\ the inheritance
diagrams of Figs.~\ref{f:vec:tensorfield_classes} and
\ref{f:man:scalar_classes} (taking into account that
\code{CommutativeAlgebraElement} is a subclass of \code{Element}).
In the present case, \code{left} = \code{v} and \code{right} = \code{f*v}
have the same parent, so that the actual result is computed in line~12,
via the method \code{\_add\_()}
(note the single underscore on each side of \code{add}). This operator is
implemented at the level of \code{TensorField}, as it can be checked from the source code
(see lines~3 and 29 below):
\begin{NBin}
v._add_??
\end{NBin}
\begin{lstlisting}
def _add_(self, other):
    """
    Tensor field addition.

    INPUT:

    - ``other`` -- a tensor field, in the same tensor module as ``self``

    OUTPUT:

    - the tensor field resulting from the addition of ``self``
      and ``other``
    ...
    """
    resu_rst = {}
    for dom in self._common_subdomains(other):
        resu_rst[dom] = self._restrictions[dom] + other._restrictions[dom]
    some_rst = next(itervalues(resu_rst))
    resu_sym = some_rst._sym
    resu_antisym = some_rst._antisym
    resu = self._vmodule.tensor(self._tensor_type, sym=resu_sym,
                                antisym=resu_antisym)
    resu._restrictions = resu_rst
    if self._name is not None and other._name is not None:
        resu._name = self._name + '+' + other._name
    if self._latex_name is not None and other._latex_name is not None:
        resu._latex_name = self._latex_name + '+' + other._latex_name
    return resu
File:  .../site-packages/sage/manifolds/differentiable/tensorfield.py
\end{lstlisting}
The first step in the addition of two vector fields is to search in the
restrictions of both vector fields for common domains: this is performed in
line~16, via the method \code{\_common\_subdomains}. Then the addition is
performed at the level of the restrictions, in line~17. The rest of the code
is simply the set up of the vector field object containing the result.
Recursively, the addition performed in line~17 will reach a level at which
the domains are parallelizable. Then a different method \code{\_add\_()}, will
be involved, as we can check on \code{vU}:
\begin{NBin}
vU._add_??
\end{NBin}
\begin{lstlisting}
def _add_(self, other):
    """
    Tensor addition.

    INPUT:

    - ``other`` -- a tensor, of the same type as ``self``

    OUTPUT:

    - the tensor resulting from the addition of ``self`` and ``other``
    ...
    """
    # No need for consistency check since self and other are guaranted
    # to belong to the same tensor module
    basis = self.common_basis(other)
    if basis is None:
        raise ValueError("no common basis for the addition")
    comp_result = self._components[basis] + other._components[basis]
    result = self._fmodule.tensor_from_comp(self._tensor_type, comp_result)
    if self._name is not None and other._name is not None:
        result._name = self._name + '+' + other._name
    if self._latex_name is not None and other._latex_name is not None:
        result._latex_name = self._latex_name + '+' + other._latex_name
    return result
File:   .../site-packages/sage/tensor/modules/free_module_tensor.py
\end{lstlisting}
From line~26, we see that this method \code{\_add\_()} is implemented
at the level of tensors on free modules, i.e.\ in the class
\code{FreeModuleTensor},\footurl{http://doc.sagemath.org/html/en/reference/tensor_free_modules/sage/tensor/modules/free_module_tensor.html}
from which \code{VectorFieldParal} inherits (cf.\ the diagram in
Fig.~\ref{f:vec:tensorfield_classes}). Here the free module is clearly
$\mathfrak{X}(U)$. The addition amounts to adding the components in a
basis of the free module in which both operands have known components. Such
a basis is returned by the method \code{common\_basis} invoked in line~16.
If necessary, this method can use change-of-basis formulas to compute the
components of \code{self} or \code{other} in a common basis.
The addition of the components in the found basis is performed in line~19. It
involves the method \code{\_\_add\_\_()} of class \code{Components}; we
can examine the corresponding code via the object \code{vUc} since the
latter has been defined above as \code{vUc = vU.\_components[eU]}, i.e.
\code{vUc} represents the set of components of the vector field
$\left. \w{v}\right| _U$ in the basis \code{eU} $=(\partial/\partial x, \partial/\partial y)$
of $\mathfrak{X}(U)$:
\begin{NBin}
vUc.__add__??
\end{NBin}
\begin{lstlisting}
def __add__(self, other):
    """
    Component addition.

    INPUT:

    - ``other`` -- components of the same number of indices and defined
      on the same frame as ``self``

    OUTPUT:

    - components resulting from the addition of ``self`` and ``other``
    ...
    """
    ...
    result = self.copy()
    nproc = Parallelism().get('tensor')
    if nproc != 1 :
        # Parallel computation
        ...
    else:
        # Sequential computation
        for ind, val in other._comp.items():
            result[[ind]] += val
    return result
File:   .../site-packages/sage/tensor/modules/comp.py
\end{lstlisting}
First of all, we note from line~26
that this is not the method \code{\_\_add\_\_()} of class \code{Element},
as it was for \code{VectorField} and \code{VectorFieldParal},
but instead the method \code{\_\_add\_\_()} implemented
in class \code{Components}. This is because \code{Components} is a
\emph{technical} class, as opposed to the \emph{mathematical} classes
\code{VectorField} and \code{DiffScalarField}; therefore it does not
inherits from \code{Element}, but only from the base class \code{SageObject},
which does not implement any addition.
We note from lines~17-19
that the computation of the components can be done in parallel on more that one CPU core if user has turned on parallelization.\footnote{This is done with the command
\code{Parallelism().set(nproc=8)} (for 8 threads); many examples of
parallelized computations are presented at
\url{https://sagemanifolds.obspm.fr/examples.html}.}
Focusing on the sequential code (lines~23-24), we see that the addition is
performed component by component.
Note that this addition is that
of scalar fields, as discussed in Sec.~\ref{s:man:add_implement},
since each component being an element of $C^\infty(U)$, the base ring of $\mathfrak{X}(U)$.

\subsection{Action of vector fields on scalar fields} \label{s:vec:action_on_scalar}

The action of $\w{v}$ on $f$ is defined pointwise by
considering $\w{v}$ at each point $p\in M$ as a derivation (the very definition of a tangent vector,
cf.\ Sec.~\ref{s:vec:def_tangent_vector}); the result is then a scalar field $\w{v}(f)$ on $M$:
\begin{NBin}
vf = v(f)
vf
\end{NBin}
\begin{NBoutM}
v\left(f\right)
\end{NBoutM}
\begin{NBin}
vf.display()
\end{NBin}
\begin{NBoutM}
\begin{array}{llcl} v\left(f\right):& M & \longrightarrow & \mathbb{R} \\ \mbox{on}\ U : & \left(x, y\right) & \longmapsto & \frac{2 \, {\left(2 \, y^{3} + 2 \, {\left(x^{2} + 1\right)} y - x\right)}}{x^{6} + y^{6} + 3 \, {\left(x^{2} + 1\right)} y^{4} + 3 \, x^{4} + 3 \, {\left(x^{4} + 2 \, x^{2} + 1\right)} y^{2} + 3 \, x^{2} + 1} \\[1ex] \mbox{on}\ V : & \left({x'}, {y'}\right) & \longmapsto & -\frac{2 \, {\left({x'}^{5} + 2 \, {x'}^{3} {y'}^{2} + {x'} {y'}^{4} - 2 \, {y'}^{5} - 2 \, {\left(2 \, {x'}^{2} + 1\right)} {y'}^{3} - 2 \, {\left({x'}^{4} + {x'}^{2}\right)} {y'}\right)}}{{x'}^{6} + {y'}^{6} + 3 \, {\left({x'}^{2} + 1\right)} {y'}^{4} + 3 \, {x'}^{4} + 3 \, {\left({x'}^{4} + 2 \, {x'}^{2} + 1\right)} {y'}^{2} + 3 \, {x'}^{2} + 1} \end{array}
\end{NBoutM}

\chapter{Tensor fields} \label{s:ten}

\minitoc

\section{Introduction}

Having presented vector fields in Chap.~\ref{s:vec}, we move now to more
general tensor fields. We keep the same example manifold, $M = \mathbb{S}^2$,
as in Chap.~\ref{s:man} and \ref{s:vec}.

\section{Differential forms}

Let us continue with the same example notebook as that considered in
Chap.~\ref{s:vec}. There, we had introduced $f$ as a scalar field on
the 2-dimensional manifold $M = \mathbb{S}^2$ (cf.\ Sec.~\ref{s:vec:tangent_impl}).
The differential of $f$ is a 1-form on $M$:
\begin{NBin}
df = f.differential()
df
\end{NBin}
\begin{NBoutM}
\mathrm{d}f
\end{NBoutM}
\begin{NBin}
print(df)
\end{NBin}
\begin{NBprint}
1-form df on the 2-dimensional differentiable manifold M
\end{NBprint}
A 1-form is actually a tensor field of type $(0,1)$:
\begin{NBin}
df.tensor_type()
\end{NBin}
\begin{NBoutM}
\left(0, 1\right)
\end{NBoutM}
while a vector field is a tensor field of type $(1,0)$:
\begin{NBin}
v.tensor_type()
\end{NBin}
\begin{NBoutM}
\left(1, 0\right)
\end{NBoutM}
Specific 1-forms are those forming the dual basis (coframe) of a given vector
frame: for instance for the vector frame \code{eU} = $(\dert{}{x},\dert{}{y})$
on $U$, considered as a basis of the free $C^\infty(U)$-module $\mathfrak{X}(U)$,
we have:
\begin{NBin}
eU.dual_basis()
\end{NBin}
\begin{NBoutM}
\left(U, \left(\mathrm{d} x,\mathrm{d} y\right)\right)
\end{NBoutM}
\begin{NBin}
print(eU.dual_basis()[0])
\end{NBin}
\begin{NBprint}
1-form dx on the Open subset U of the 2-dimensional differentiable manifold M
\end{NBprint}
Since \code{eU} is the default frame on $M$, the default display of $\mathrm{d}f$
is performed in terms of \code{eU}'s coframe:
\begin{NBin}
df.display()
\end{NBin}
\begin{NBoutM}
\mathrm{d}f = \left( -\frac{2 \, x}{x^{4} + y^{4} + 2 \, {\left(x^{2} + 1\right)} y^{2} + 2 \, x^{2} + 1} \right) \mathrm{d} x + \left( -\frac{2 \, y}{x^{4} + y^{4} + 2 \, {\left(x^{2} + 1\right)} y^{2} + 2 \, x^{2} + 1} \right) \mathrm{d} y
\end{NBoutM}
We may check that in this basis, the components of $\left. \mathrm{d}f \right| _U$
are nothing but the partial derivatives of the coordinate expression of $f$
with respect to coordinates $(x,y)$:
\begin{NBin}
df[0] == diff(f.expr(), x)
\end{NBin}
\begin{NBout}
\texttt{True}
\end{NBout}
\begin{NBin}
df[1] == diff(f.expr(), y)
\end{NBin}
\begin{NBout}
\texttt{True}
\end{NBout}
In the coframe associated with \code{eV} = $(\dert{}{x'},\dert{}{y'})$:
\begin{NBin}
df.display(eV)
\end{NBin}
\begin{NBout}
$\displaystyle
\mathrm{d}f = \left( \frac{2 \, {x'}}{{x'}^{4} + {y'}^{4} + 2 \, {\left({x'}^{2} + 1\right)} {y'}^{2} + 2 \, {x'}^{2} + 1} \right) \mathrm{d} {x'} $\\
$\displaystyle
+ \left( \frac{2 \, {y'}}{{x'}^{4} + {y'}^{4} + 2 \, {\left({x'}^{2} + 1\right)} {y'}^{2} + 2 \, {x'}^{2} + 1} \right) \mathrm{d} {y'}$
\end{NBout}
Since \code{eV} is not the default vector frame on $M$ and \code{XV} = $(V,(x',y'))$
is not the default chart on $M$, we get the individual components by
specifying both \code{eV} and \code{XV}, in addition to the index, in the
square-bracket operator:
\begin{NBin}
df[eV,0,XV]
\end{NBin}
\begin{NBoutM}
\frac{2 \, {x'}}{{x'}^{4} + {y'}^{4} + 2 \, {\left({x'}^{2} + 1\right)} {y'}^{2} + 2 \, {x'}^{2} + 1}
\end{NBoutM}
We may then check that the components in the frame \code{eV}
are the partial derivatives with respect to the coordinates \code{xp} = $x'$ and
\code{yp} = $y'$ of the chart \code{XV}:
\begin{NBin}
df[eV,0,XV] == diff(f.expr(XV), xp)
\end{NBin}
\begin{NBout}
\texttt{True}
\end{NBout}
\begin{NBin}
df[eV,1,XV] == diff(f.expr(XV), yp)
\end{NBin}
\begin{NBout}
\texttt{True}
\end{NBout}
The parent of $\mathrm{d}f$ is the set $\Omega^1(M)$ of all 1-forms on $M$,
considered as a $C^\infty(M)$-module:
\begin{NBin}
print(df.parent())
df.parent()
\end{NBin}
\begin{NBprint}
Module Omega^1(M) of 1-forms on the 2-dimensional differentiable manifold M
\end{NBprint}
\begin{NBoutM}
\Omega^{1}\left(M\right)
\end{NBoutM}
\begin{NBin}
df.parent().base_ring()
\end{NBin}
\begin{NBoutM}
C^{\infty}\left(M\right)
\end{NBoutM}
This module is actually the dual of the vector-field module $\mathfrak{X}(M)$,
which is represented
by the Python object \code{YM} (cf.\ Sec.~\ref{s:vec:vector_field_impl}):
\begin{NBin}
YM.dual()
\end{NBin}
\begin{NBoutM}
\Omega^{1}\left(M\right)
\end{NBoutM}
Consequently, a 1-form acts on vector fields, yielding an element of
$C^\infty(M)$, i.e.\ a scalar field:
\begin{NBin}
print(df(v))
\end{NBin}
\begin{NBprint}
Scalar field df(v) on the 2-dimensional differentiable manifold M
\end{NBprint}
This scalar field is nothing but the result of the action of $\w{v}$ on $f$
discussed in Sec.~\ref{s:vec:action_on_scalar}:
\begin{NBin}
df(v) == v(f)
\end{NBin}
\begin{NBout}
\texttt{True}
\end{NBout}

\section{More general tensor fields}

We construct a tensor of type $(1,1)$ by taking the tensor product
$\w{v}\otimes \mathrm{d}f$:
\begin{NBin}
t = v * df
t
\end{NBin}
\begin{NBoutM}
\mbox{Tensor field of type (1,1) on the 2-dimensional differentiable manifold M}
\end{NBoutM}
\begin{NBin}
t.display()
\end{NBin}
\begin{NBoutM}
v\otimes \mathrm{d}f = \left( -\frac{2 \, x}{x^{6} + y^{6} + 3 \, {\left(x^{2} + 1\right)} y^{4} + 3 \, x^{4} + 3 \, {\left(x^{4} + 2 \, x^{2} + 1\right)} y^{2} + 3 \, x^{2} + 1} \right) \frac{\partial}{\partial x }\otimes \mathrm{d} x + \left( -\frac{2 \, y}{x^{6} + y^{6} + 3 \, {\left(x^{2} + 1\right)} y^{4} + 3 \, x^{4} + 3 \, {\left(x^{4} + 2 \, x^{2} + 1\right)} y^{2} + 3 \, x^{2} + 1} \right) \frac{\partial}{\partial x }\otimes \mathrm{d} y + \left( \frac{4 \, x}{x^{4} + y^{4} + 2 \, {\left(x^{2} + 1\right)} y^{2} + 2 \, x^{2} + 1} \right) \frac{\partial}{\partial y }\otimes \mathrm{d} x + \left( \frac{4 \, y}{x^{4} + y^{4} + 2 \, {\left(x^{2} + 1\right)} y^{2} + 2 \, x^{2} + 1} \right) \frac{\partial}{\partial y }\otimes \mathrm{d} y
\end{NBoutM}
\begin{NBin}
t.display(eV)
\end{NBin}
\begin{NBoutM}
v\otimes \mathrm{d}f = \left( -\frac{2 \, {\left({x'}^{5} - 4 \, {x'}^{2} {y'}^{3} - {x'} {y'}^{4} - 4 \, {\left({x'}^{4} + {x'}^{2}\right)} {y'}\right)}}{{x'}^{6} + {y'}^{6} + 3 \, {\left({x'}^{2} + 1\right)} {y'}^{4} + 3 \, {x'}^{4} + 3 \, {\left({x'}^{4} + 2 \, {x'}^{2} + 1\right)} {y'}^{2} + 3 \, {x'}^{2} + 1} \right) \frac{\partial}{\partial {x'} }\otimes \mathrm{d} {x'} + \left( -\frac{2 \, {\left({x'}^{4} {y'} - 4 \, {x'} {y'}^{4} - {y'}^{5} - 4 \, {\left({x'}^{3} + {x'}\right)} {y'}^{2}\right)}}{{x'}^{6} + {y'}^{6} + 3 \, {\left({x'}^{2} + 1\right)} {y'}^{4} + 3 \, {x'}^{4} + 3 \, {\left({x'}^{4} + 2 \, {x'}^{2} + 1\right)} {y'}^{2} + 3 \, {x'}^{2} + 1} \right) \frac{\partial}{\partial {x'} }\otimes \mathrm{d} {y'} + \left( -\frac{4 \, {\left({x'}^{5} + {x'}^{4} {y'} + {x'}^{2} {y'}^{3} - {x'} {y'}^{4} + {x'}^{3} - {x'} {y'}^{2}\right)}}{{x'}^{6} + {y'}^{6} + 3 \, {\left({x'}^{2} + 1\right)} {y'}^{4} + 3 \, {x'}^{4} + 3 \, {\left({x'}^{4} + 2 \, {x'}^{2} + 1\right)} {y'}^{2} + 3 \, {x'}^{2} + 1} \right) \frac{\partial}{\partial {y'} }\otimes \mathrm{d} {x'} + \left( -\frac{4 \, {\left({x'}^{3} {y'}^{2} + {x'} {y'}^{4} - {y'}^{5} - {y'}^{3} + {\left({x'}^{4} + {x'}^{2}\right)} {y'}\right)}}{{x'}^{6} + {y'}^{6} + 3 \, {\left({x'}^{2} + 1\right)} {y'}^{4} + 3 \, {x'}^{4} + 3 \, {\left({x'}^{4} + 2 \, {x'}^{2} + 1\right)} {y'}^{2} + 3 \, {x'}^{2} + 1} \right) \frac{\partial}{\partial {y'} }\otimes \mathrm{d} {y'}
\end{NBoutM}
We can use the method \code{display\_comp()} for a display component by
component:
\begin{NBin}
t.display_comp()
\end{NBin}
\begin{NBoutM}
\begin{array}{lcl} v\otimes \mathrm{d}f_{ \phantom{\, x} \, x }^{ \, x \phantom{\, x} } & = & -\frac{2 \, x}{x^{6} + y^{6} + 3 \, {\left(x^{2} + 1\right)} y^{4} + 3 \, x^{4} + 3 \, {\left(x^{4} + 2 \, x^{2} + 1\right)} y^{2} + 3 \, x^{2} + 1} \\ v\otimes \mathrm{d}f_{ \phantom{\, x} \, y }^{ \, x \phantom{\, y} } & = & -\frac{2 \, y}{x^{6} + y^{6} + 3 \, {\left(x^{2} + 1\right)} y^{4} + 3 \, x^{4} + 3 \, {\left(x^{4} + 2 \, x^{2} + 1\right)} y^{2} + 3 \, x^{2} + 1} \\ v\otimes \mathrm{d}f_{ \phantom{\, y} \, x }^{ \, y \phantom{\, x} } & = & \frac{4 \, x}{x^{4} + y^{4} + 2 \, {\left(x^{2} + 1\right)} y^{2} + 2 \, x^{2} + 1} \\ v\otimes \mathrm{d}f_{ \phantom{\, y} \, y }^{ \, y \phantom{\, y} } & = & \frac{4 \, y}{x^{4} + y^{4} + 2 \, {\left(x^{2} + 1\right)} y^{2} + 2 \, x^{2} + 1} \end{array}
\end{NBoutM}
The parent of $t$ is the set $\mathcal{T}^{(1,1)}(M)$ of all type-$(1,1)$
tensor fields on $M$,
considered as a $C^\infty(M)$-module:
\begin{NBin}
print(t.parent())
t.parent()
\end{NBin}
\begin{NBprint}
Module T^(1,1)(M) of type-(1,1) tensors fields on the 2-dimensional
differentiable manifold M
\end{NBprint}
\begin{NBoutM}
\mathcal{T}^{(1,1)}\left(M\right)
\end{NBoutM}
\begin{NBin}
t.parent().base_ring()
\end{NBin}
\begin{NBoutM}
C^{\infty}\left(M\right)
\end{NBoutM}

As for vector fields, since $M$ is not parallelizable, the $C^\infty(M)$-module
$\mathcal{T}^{(1,1)}(M)$ is not free and the tensor fields are described by
their restrictions to parallelizable subdomains:
\begin{NBin}
t._restrictions
\end{NBin}
\begin{NBoutM}
\left\{V : v\otimes \mathrm{d}f, U : v\otimes \mathrm{d}f\right\}
\end{NBoutM}
These restrictions form free modules:
\begin{NBin}
print(t._restrictions[U].parent())
\end{NBin}
\begin{NBprint}
Free module T^(1,1)(U) of type-(1,1) tensors fields on the Open subset U of
the 2-dimensional differentiable manifold M
\end{NBprint}
\begin{NBin}
t._restrictions[U].parent().base_ring()
\end{NBin}
\begin{NBoutM}
C^{\infty}\left(U\right)
\end{NBoutM}

\section{Riemannian metric}

\subsection{Defining a metric}

The standard metric on $M=\mathbb{S}^2$ is that induced by the Euclidean metric of $\mathbb{R}^3$. Let us start by defining the latter:
\begin{NBin}
h = R3.metric('h')
h[0,0], h[1,1], h[2, 2] = 1, 1, 1
h.display()
\end{NBin}
\begin{NBoutM}
h = \mathrm{d} X\otimes \mathrm{d} X+\mathrm{d} Y\otimes \mathrm{d} Y+\mathrm{d} Z\otimes \mathrm{d} Z
\end{NBoutM}
The metric $g$ on $M$ is the pullback of $h$ associated with the embedding $\Phi$
introduced in Sec.~\ref{s:vec:tangent_impl}:
\begin{NBin}
g = M.metric('g')
g.set( Phi.pullback(h) )
print(g)
\end{NBin}
\begin{NBprint}
Riemannian metric g on the 2-dimensional differentiable manifold M
\end{NBprint}
Note that we could have defined $g$ intrinsically, i.e.\ by providing its components in the two vector frames \code{eU} and \code{eV}, as we did for the metric $h$ on $\mathbb{R}^3$. Instead, we have chosen to get it as the pullback by $\Phi$ of $h$, as an example of pullback associated with some differential map.

The metric is a symmetric tensor field of type (0,2):
\begin{NBin}
g.tensor_type()
\end{NBin}
\begin{NBoutM}
\left(0, 2\right)
\end{NBoutM}
The expression of the metric in terms of the default frame on $M$ (\code{eU}):
\begin{NBin}
g.display()
\end{NBin}
\begin{NBout}
$\displaystyle
g = \left( \frac{4}{x^{4} + y^{4} + 2 \, {\left(x^{2} + 1\right)} y^{2} + 2 \, x^{2} + 1} \right) \mathrm{d} x\otimes \mathrm{d} x $\\
$\displaystyle
+ \left( \frac{4}{x^{4} + y^{4} + 2 \, {\left(x^{2} + 1\right)} y^{2} + 2 \, x^{2} + 1} \right) \mathrm{d} y\otimes \mathrm{d} y$
\end{NBout}
We may factorize the metric components to get a better display:
\begin{NBin}
g[0,0].factor(); g[1,1].factor()
\end{NBin}
\begin{NBoutM}
\frac{4}{{\left(x^{2} + y^{2} + 1\right)}^{2}}
\end{NBoutM}
\begin{NBin}
g.display()
\end{NBin}
\begin{NBoutM}
g = \frac{4}{{\left(x^{2} + y^{2} + 1\right)}^{2}} \mathrm{d} x\otimes \mathrm{d} x + \frac{4}{{\left(x^{2} + y^{2} + 1\right)}^{2}} \mathrm{d} y\otimes \mathrm{d} y
\end{NBoutM}
A matrix view of the components of $g$ in the manifold's default frame:
\begin{NBin}
g[:]
\end{NBin}
\begin{NBoutM}
\left(\begin{array}{rr}
    \frac{4}{{\left(x^{2} + y^{2} + 1\right)}^{2}} & 0 \\
    0 & \frac{4}{{\left(x^{2} + y^{2} + 1\right)}^{2}}
    \end{array}\right)
\end{NBoutM}
Display in terms of the vector frame $(V, (\partial_{x'}, \partial_{y'}))$:
\begin{NBin}
g.display(eV)
\end{NBin}
\begin{NBout}
$\displaystyle
g = \left( \frac{4}{{x'}^{4} + {y'}^{4} + 2 \, {\left({x'}^{2} + 1\right)} {y'}^{2} + 2 \, {x'}^{2} + 1} \right) \mathrm{d} {x'}\otimes \mathrm{d} {x'}$\\
$\displaystyle + \left( \frac{4}{{x'}^{4} + {y'}^{4} + 2 \, {\left({x'}^{2} + 1\right)} {y'}^{2} + 2 \, {x'}^{2} + 1} \right) \mathrm{d} {y'}\otimes \mathrm{d} {y'}$
\end{NBout}
The metric acts on vector field pairs, resulting in a scalar field:
\begin{NBin}
print(g(v,v))
\end{NBin}
\begin{NBprint}
Scalar field g(v,v) on the 2-dimensional differentiable manifold M
\end{NBprint}
\begin{NBin}
g(v,v).parent()
\end{NBin}
\begin{NBoutM}
C^{\infty}\left(M\right)
\end{NBoutM}
\begin{NBin}
g(v,v).display()
\end{NBin}
\begin{NBoutM}
\begin{array}{llcl} g\left(v,v\right):& M & \longrightarrow & \mathbb{R} \\ \mbox{on}\ U : & \left(x, y\right) & \longmapsto & \frac{4 \, {\left(4 \, x^{4} + 4 \, y^{4} + 8 \, {\left(x^{2} + 1\right)} y^{2} + 8 \, x^{2} + 5\right)}}{x^{8} + y^{8} + 4 \, {\left(x^{2} + 1\right)} y^{6} + 4 \, x^{6} + 6 \, {\left(x^{4} + 2 \, x^{2} + 1\right)} y^{4} + 6 \, x^{4} + 4 \, {\left(x^{6} + 3 \, x^{4} + 3 \, x^{2} + 1\right)} y^{2} + 4 \, x^{2} + 1} \\[1ex] \mbox{on}\ V : & \left({x'}, {y'}\right) & \longmapsto & \frac{4 \, {\left(5 \, {x'}^{8} + 5 \, {y'}^{8} + 4 \, {\left(5 \, {x'}^{2} + 2\right)} {y'}^{6} + 8 \, {x'}^{6} + 2 \, {\left(15 \, {x'}^{4} + 12 \, {x'}^{2} + 2\right)} {y'}^{4} + 4 \, {x'}^{4} + 4 \, {\left(5 \, {x'}^{6} + 6 \, {x'}^{4} + 2 \, {x'}^{2}\right)} {y'}^{2}\right)}}{{x'}^{8} + {y'}^{8} + 4 \, {\left({x'}^{2} + 1\right)} {y'}^{6} + 4 \, {x'}^{6} + 6 \, {\left({x'}^{4} + 2 \, {x'}^{2} + 1\right)} {y'}^{4} + 6 \, {x'}^{4} + 4 \, {\left({x'}^{6} + 3 \, {x'}^{4} + 3 \, {x'}^{2} + 1\right)} {y'}^{2} + 4 \, {x'}^{2} + 1} \end{array}
\end{NBoutM}

\subsection{Levi-Civita connection}

The Levi-Civita connection associated with the metric $g$ is
\begin{NBin}
nab = g.connection()
print(nab)
nab
\end{NBin}
\begin{NBprint}
Levi-Civita connection nabla_g associated with the Riemannian metric g on
the 2-dimensional differentiable manifold M
\end{NBprint}
\begin{NBoutM}
\nabla_{g}
\end{NBoutM}
The nonzero Christoffel symbols of $g$ (skipping those that can be deduced by symmetry on the last two indices) w.r.t. the chart \code{XU}:
\begin{NBin}
g.christoffel_symbols_display(chart=XU)
\end{NBin}
\begin{NBoutM}
\begin{array}{lcl} \Gamma_{ \phantom{\, x} \, x \, x }^{ \, x \phantom{\, x} \phantom{\, x} } & = & -\frac{2 \, x}{x^{2} + y^{2} + 1} \\ \Gamma_{ \phantom{\, x} \, x \, y }^{ \, x \phantom{\, x} \phantom{\, y} } & = & -\frac{2 \, y}{x^{2} + y^{2} + 1} \\ \Gamma_{ \phantom{\, x} \, y \, y }^{ \, x \phantom{\, y} \phantom{\, y} } & = & \frac{2 \, x}{x^{2} + y^{2} + 1} \\ \Gamma_{ \phantom{\, y} \, x \, x }^{ \, y \phantom{\, x} \phantom{\, x} } & = & \frac{2 \, y}{x^{2} + y^{2} + 1} \\ \Gamma_{ \phantom{\, y} \, x \, y }^{ \, y \phantom{\, x} \phantom{\, y} } & = & -\frac{2 \, x}{x^{2} + y^{2} + 1} \\ \Gamma_{ \phantom{\, y} \, y \, y }^{ \, y \phantom{\, y} \phantom{\, y} } & = & -\frac{2 \, y}{x^{2} + y^{2} + 1} \end{array}
\end{NBoutM}
$\nabla_g$ acting on the vector field $\w{v}$:
\begin{NBin}
Dv = nab(v)
print(Dv)
\end{NBin}
\begin{NBprint}
Tensor field nabla_g(v) of type (1,1) on the 2-dimensional differentiable
manifold M
\end{NBprint}
\begin{NBin}
Dv.display()
\end{NBin}
\begin{NBout}
$\displaystyle
\nabla_{g} v = \left( \frac{4 \, {\left(y^{3} + {\left(x^{2} + 1\right)} y - x\right)}}{x^{4} + y^{4} + 2 \, {\left(x^{2} + 1\right)} y^{2} + 2 \, x^{2} + 1} \right) \frac{\partial}{\partial x }\otimes \mathrm{d} x $\\
$\displaystyle
+ \left( -\frac{4 \, {\left(x^{3} + x y^{2} + x + y\right)}}{x^{4} + y^{4} + 2 \, {\left(x^{2} + 1\right)} y^{2} + 2 \, x^{2} + 1} \right) \frac{\partial}{\partial x }\otimes \mathrm{d} y $\\
$\displaystyle
+ \left( \frac{2 \, {\left(2 \, x^{3} + 2 \, x y^{2} + 2 \, x + y\right)}}{x^{4} + y^{4} + 2 \, {\left(x^{2} + 1\right)} y^{2} + 2 \, x^{2} + 1} \right) \frac{\partial}{\partial y }\otimes \mathrm{d} x$\\
$\displaystyle
 + \left( \frac{2 \, {\left(2 \, y^{3} + 2 \, {\left(x^{2} + 1\right)} y - x\right)}}{x^{4} + y^{4} + 2 \, {\left(x^{2} + 1\right)} y^{2} + 2 \, x^{2} + 1} \right) \frac{\partial}{\partial y }\otimes \mathrm{d} y$
\end{NBout}

\subsection{Curvature}

The Riemann curvature tensor of the metric $g$ is
\begin{NBin}
Riem = g.riemann()
print(Riem)
Riem.display()
\end{NBin}
\begin{NBprint}
Tensor field Riem(g) of type (1,3) on the 2-dimensional differentiable
manifold M
\end{NBprint}
\begin{NBout}
$\displaystyle
\mathrm{Riem}\left(g\right) = \left( \frac{4}{x^{4} + y^{4} + 2 \, {\left(x^{2} + 1\right)} y^{2} + 2 \, x^{2} + 1} \right) \frac{\partial}{\partial x }\otimes \mathrm{d} y\otimes \mathrm{d} x\otimes \mathrm{d} y $\\
$\displaystyle + \left( -\frac{4}{x^{4} + y^{4} + 2 \, {\left(x^{2} + 1\right)} y^{2} + 2 \, x^{2} + 1} \right) \frac{\partial}{\partial x }\otimes \mathrm{d} y\otimes \mathrm{d} y\otimes \mathrm{d} x$\\
$\displaystyle
 + \left( -\frac{4}{x^{4} + y^{4} + 2 \, {\left(x^{2} + 1\right)} y^{2} + 2 \, x^{2} + 1} \right) \frac{\partial}{\partial y }\otimes \mathrm{d} x\otimes \mathrm{d} x\otimes \mathrm{d} y$\\
$\displaystyle
 + \left( \frac{4}{x^{4} + y^{4} + 2 \, {\left(x^{2} + 1\right)} y^{2} + 2 \, x^{2} + 1} \right) \frac{\partial}{\partial y }\otimes \mathrm{d} x\otimes \mathrm{d} y\otimes \mathrm{d} x$
\end{NBout}
The components of the Riemann tensor in the default frame on $M$ are
\begin{NBin}
Riem.display_comp()
\end{NBin}
\begin{NBoutM}
\begin{array}{lcl} \mathrm{Riem}\left(g\right)_{ \phantom{\, x} \, y \, x \, y }^{ \, x \phantom{\, y} \phantom{\, x} \phantom{\, y} } & = & \frac{4}{x^{4} + y^{4} + 2 \, {\left(x^{2} + 1\right)} y^{2} + 2 \, x^{2} + 1} \\ \mathrm{Riem}\left(g\right)_{ \phantom{\, x} \, y \, y \, x }^{ \, x \phantom{\, y} \phantom{\, y} \phantom{\, x} } & = & -\frac{4}{x^{4} + y^{4} + 2 \, {\left(x^{2} + 1\right)} y^{2} + 2 \, x^{2} + 1} \\ \mathrm{Riem}\left(g\right)_{ \phantom{\, y} \, x \, x \, y }^{ \, y \phantom{\, x} \phantom{\, x} \phantom{\, y} } & = & -\frac{4}{x^{4} + y^{4} + 2 \, {\left(x^{2} + 1\right)} y^{2} + 2 \, x^{2} + 1} \\ \mathrm{Riem}\left(g\right)_{ \phantom{\, y} \, x \, y \, x }^{ \, y \phantom{\, x} \phantom{\, y} \phantom{\, x} } & = & \frac{4}{x^{4} + y^{4} + 2 \, {\left(x^{2} + 1\right)} y^{2} + 2 \, x^{2} + 1} \end{array}
\end{NBoutM}
The parent of the Riemann tensor is the $C^\infty(M)$-module of
type-(1,3) tensor fields on $M$:
\begin{NBin}
print(Riem.parent())
\end{NBin}
\begin{NBprint}
Module T^(1,3)(M) of type-(1,3) tensors fields on the 2-dimensional
differentiable manifold M
\end{NBprint}
The Riemann tensor is antisymmetric on its two last indices (i.e.\ the indices
at position 2 and 3, the first index being at position 0):
\begin{NBin}
Riem.symmetries()
\end{NBin}
\begin{NBout}
\texttt{no symmetry; antisymmetry:~(2, 3)}
\end{NBout}
The Riemann tensor of the Euclidean metric $h$ on $\mathbb{R}^3$ is identically zero,
i.e.\ $h$ is a flat metric:
\begin{NBin}
h.riemann().display()
\end{NBin}
\begin{NBoutM}
\mathrm{Riem}\left(h\right) = 0
\end{NBoutM}
The Ricci tensor is
\begin{NBin}
Ric = g.ricci()
Ric.display()
\end{NBin}
\begin{NBout}
$\displaystyle
\mathrm{Ric}\left(g\right) = \left( \frac{4}{x^{4} + y^{4} + 2 \, {\left(x^{2} + 1\right)} y^{2} + 2 \, x^{2} + 1} \right) \mathrm{d} x\otimes \mathrm{d} x $\\
$\displaystyle
+ \left( \frac{4}{x^{4} + y^{4} + 2 \, {\left(x^{2} + 1\right)} y^{2} + 2 \, x^{2} + 1} \right) \mathrm{d} y\otimes \mathrm{d} y$
\end{NBout}
while the Ricci scalar is
\begin{NBin}
R = g.ricci_scalar()
R.display()
\end{NBin}
\begin{NBoutM}
\begin{array}{llcl} \mathrm{r}\left(g\right):& M & \longrightarrow & \mathbb{R} \\ \mbox{on}\ U : & \left(x, y\right) & \longmapsto & 2 \\ \mbox{on}\ V : & \left({x'}, {y'}\right) & \longmapsto & 2 \end{array}
\end{NBoutM}
We recover the fact that $(\mathbb{S}^2,g)$ is a Riemannian manifold of constant positive curvature.

In dimension 2, the Riemann curvature tensor is entirely determined by the Ricci scalar $R$ according to
\be
 R^i_{\ \, jlk} = \frac{R}{2} \left( \delta^i_{\ \, k} g_{jl} - \delta^i_{\ \, l} g_{jk} \right)
\ee
Let us check this formula here, under the form
$R^i_{\ \, jlk} = -R g_{j[k} \delta^i_{\ \, l]}$:
\begin{NBin}
delta = M.tangent_identity_field()
Riem == - R*(g*delta).antisymmetrize(2,3)
\end{NBin}
\begin{NBout}
\texttt{True}
\end{NBout}
Similarly the relation $\mathrm{Ric} = (R/2)\; g$ must hold:
\begin{NBin}
Ric == (R/2)*g
\end{NBin}
\begin{NBout}
\texttt{True}
\end{NBout}

\subsection{Volume form}

The \defin{volume form} (or \defin{Levi-Civita tensor}) associated with the
metric $g$ and for which the vector frame $(\partial_x,\partial_y)$ is
right-handed is the following 2-form:
\begin{NBin}
eps = g.volume_form()
print(eps)
eps.display()
\end{NBin}
\begin{NBoutM}
\epsilon_{g} = \left( \frac{4}{x^{4} + y^{4} + 2 \, {\left(x^{2} + 1\right)} y^{2} + 2 \, x^{2} + 1} \right) \mathrm{d} x\wedge \mathrm{d} y
\end{NBoutM}
The exterior derivative of $\epsilon_g$ is a 3-form:
\begin{NBin}
print(eps.exterior_derivative())
\end{NBin}
\begin{NBprint}
3-form deps_g on the 2-dimensional differentiable manifold M
\end{NBprint}
Of course, since the dimension of $M$ is 2, all 3-forms vanish identically:
\begin{NBin}
eps.exterior_derivative().display()
\end{NBin}
\begin{NBoutM}
\mathrm{d}\epsilon_{g} = 0
\end{NBoutM}

\chapter{Conclusion and perspectives} \label{s:con}

We have presented some aspects of symbolic tensor calculus as implemented
in \Sage{}. The implementation is independent of the symbolic backend (i.e. the
tool used to performed symbolic calculus on coordinate representations of
scalar fields), the latter being involved only in the last stage of the diagram
shown in Fig.~\ref{f:vec:storage_tensor}.

The implementation has been performed via the \soft{SageManifolds}
project, the home page of which we refer to for details and material complementary
to what has been shown here (in particular many more examples):
\begin{center}
\url{https://sagemanifolds.obspm.fr/}
\end{center}
This project resulted in approximately $85,000$ lines of Python code (including comments and doctests), which have been submitted to \Sage{} community as a sequence of
$\sim 50$ tickets\footnote{Cf.\ the meta-ticket \url{https://trac.sagemath.org/ticket/18528}.}
at the time of this writing (October 2018), the
first ticket having been accepted in March 2015.
These tickets have been written and reviewed by a dozen of
contributors.\footnote{Cf.\ the list at \url{https://sagemanifolds.obspm.fr/authors.html}.}
As a result, all code is fully included in \soft{SageMath~8.4} and does not require
any separate installation. The following features have been already implemented:
\begin{itemize}
\item differentiable manifolds: tangent spaces, vector frames, tensor fields, curves, pullback and pushforward operators;
\item standard tensor calculus (tensor product, contraction, symmetrization, etc.), even on non-parallelizable manifolds;
\item all monoterm tensor symmetries taken into account;
\item Lie derivatives of tensor fields;
\item differential forms: exterior and interior products, exterior derivative,
Hodge duality;
\item multivector fields: exterior and interior products, Schouten-Nijenhuis bracket;
\item affine connections (curvature, torsion);
\item pseudo-Riemannian metrics;
\item computation of geodesics (numerical integration via \Sage{}/\soft{GSL});
\item some plotting capabilities (charts, points, curves, vector fields);
\item extrinsic geometry of pseudo-Riemannian submanifolds;
\item parallelization (on tensor components) of CPU demanding computations,
via the Python library \code{multiprocessing};
\item the possibility to use \soft{SymPy} as the symbolic backend, instead of
\Sage{}'s default, which is \soft{Pynac} (with \soft{Maxima} for simplifications).
\end{itemize}
Only a subset of the above functionalities have been presented in these lectures.
In particular,
the exterior calculus on differential forms and multivector fields has not been
touched, nor the computation of geodesics.

\medskip

The \soft{SageManifolds} project is still ongoing and future prospects include
\begin{itemize}
\item adding more symbolic backends (\soft{Giac}, \soft{FriCAS}, ...);
\item computing integrals on submanifolds;
\item adding more plotting capabilities;
\item introducing new functionalities: symplectic forms, fibre bundles,
spinors, variational calculus, etc.;
\item connecting with numerical relativity: using \Sage{} to explore
numerically-generated spacetimes; this will be done by introducing
\emph{numerical} backends, instead
of \emph{symbolic} ones, in the last stage of the Fig.~\ref{f:vec:storage_tensor}
diagram.
\end{itemize}
In the spirit of open-source software, anybody interested is very welcome
to join the project. Please visit
\begin{center}
\url{https://sagemanifolds.obspm.fr/contact.html}
\end{center}


\end{document}